\documentclass[lettersize,journal]{IEEEtran}
\usepackage{setspace}
\usepackage{amsmath,amsfonts}
\usepackage{algorithmic}
\usepackage{algorithm}
\usepackage{array}
\usepackage[caption=false,font=normalsize,labelfont=sf,textfont=sf]{subfig}
\usepackage{textcomp}
\usepackage{stfloats}
\usepackage{url}
\usepackage{verbatim}
\usepackage{graphicx}
\usepackage{cite}
\usepackage{xcolor}
\usepackage{amssymb}
\usepackage{booktabs}
\usepackage{multirow}
\usepackage{subfig}
\usepackage{float}

\graphicspath{{fig/}}

\begin{document}

\title{Electromagnetic Information Theory: Fundamentals, Paradigm Shifts, and Applications}

\author{
	Tengjiao Wang, Zhenyu Kang, Ting Li, Zhihui Chen, Shaobo Wang,
	
	Yingpei Lin, Yan Wang, and Yichuan Yu

	\thanks{\textit{Corresponding authors: Shaobo Wang.}}
	\thanks{T. Wang, Z. Kang, T. Li, Z. Chen, S. Wang, Y. Lin, Y. Wang, Y. Yu are with the Department of Wireless Network Research, Huawei Technologies CO., Ltd, Shanghai 201206, China (e-mail: \{wangtengjiao6, kangzhenyu, liting147, chenzhihui28, shaobo.wang, linyingpei, eleanor.wangyan, yuyichuan\}@huawei.com).}

}

\maketitle

\begin{abstract}	
This paper explores the emerging research direction of electromagnetic information theory (EIT), which aims to integrate traditional Shannon-based methodologies with physical consistency, particularly the electromagnetic properties of communication channels. We propose an EIT-based multiple-input multiple-output (MIMO) paradigm that enhances conventional spatially-discrete MIMO models by incorporating the concepts of electromagnetic (EM) precoding and EM combining. This approach aims to improve the modeling of next-generation systems while remaining consistent with Shannon's theoretical foundations. We explore typical EIT applications, such as densely spaced MIMO, near-field communications, and tri-polarized antennas, and analyze their channel characteristics through theoretical simulations and measured datasets. The paper also discusses critical research challenges and opportunities for EIT applications from an industrial perspective, emphasizing the field's potential for practical applications.
\end{abstract}



\vspace{-10pt}
\section{Introduction}

The wireless communication landscape is rapidly evolving with the progression from fifth-generation (5G) to 5G-advanced technologies. This evolution has brought innovative application scenarios, including the Internet of Things (IoT), autonomous vehicles, and extended reality (XR), that are redefining the boundaries of connectivity and data exchange \cite{10054381}. 
These novel applications impose stringent requirements on communication systems, such as ultra-low latency, ultra-high reliability, and large connection densities. 
In response to these demands, researchers and engineers are exploring new methodologies that integrate Shannon-based information theory with physical consistency, particularly the electromagnetic properties of communication channels. This has led to the development of electromagnetic information theory (EIT) \cite{zhu2023electromagneticinformationtheoryfundamentals,wang2024electromagneticinformationtheoryfundamentals,direnzo2023electromagneticsignalinformationtheory}, which aims to extend traditional approaches while maintaining alignment with Shannon’s foundational principles.

EIT is an emerging interdisciplinary field that seeks to integrate the principles of information theory with the physical laws of electromagnetism to better understand and optimize wireless communication systems. Building on foundational work \cite{Loyka2004,4636839,Franceschetti2018}, which sought to reconcile Shannon's theoretical limits with Maxwell's equations, EIT emphasizes the importance of understanding the electromagnetic constraints in massive multiple-input multiple-output (MIMO) communications. Rather than superseding Shannon theory, EIT aims to extend it by incorporating physical insights into electromagnetic wave propagation, which can inform and refine our understanding of information transmission and system performance in real-world wireless environments.
\cite{29386} introduced the degrees of freedom of scattered fields and provided an optimal sampling representation.
\cite{6880934} expanded on this by discussing the unavoidable limitations that the laws of electromagnetism impose on communication systems and the critical role of the degrees of freedom in radiating systems. These degrees of freedom (DoFs) play a pivotal role in determining the capabilities of radiating systems, which is a key aspect of EIT's approach to understanding antenna performance and system design. It revealed that different classes of antennas utilize these DoFs in various ways, impacting the overall performance and efficiency of wireless communications.
In addition to DoFs, entropy and capacity calculations based on EM theory are also essential concepts of EIT. \cite{4636839} discussed the usage of entropy for antenna synthesis. \cite{2017JSP...169..374F} considered the physical limits of entropy carried by band-limited waveforms. \cite{Cui2016} introduced  entropy in the context of meta-surfaces for electromagnetic applications. And in \cite{5165183}, detailed derivations of wireless network capacity using information-theoretic approaches was provided. Besides, 
\cite{8585146} further emphasized the significance of electromagnetism in wireless transmission, suggesting that the physical properties of the electromagnetic field are more critical than the information processing techniques in determining the upper limits of reliable transmission. This perspective emphasized the need to understand the underlying physics of electromagnetic fields to optimize wireless communication channels.
Building on these insights, recent works \cite{zhu2023electromagneticinformationtheoryfundamentals} and \cite{wang2024electromagneticinformationtheoryfundamentals} introduced the innovative electromagnetic (EM) physical layer concept and provided a systematic exploration of EIT's foundational principles. It facilitated a deeper understanding of the interplay between electromagnetic phenomena and information transmission, and discussed how EIT can guide the practical design of wireless systems by considering the continuous field modeling, degrees of freedom, and mutual information.
In summary, EIT aims to investigate and leverage the electromagnetic phenomena in information transmission, thereby providing a more comprehensive framework for the design and optimization of wireless communication systems. This integration is vital for pushing the boundaries of what is possible in wireless communications, thus inspiring numerous innovative applications.

First, EIT has emerged as a pivotal framework for establishing EM-compliant channel models, inspiring specific technologies to utilize the EM characteristics more efficiently. 
The foundational principles of EIT trace back to Gallager’s waveform channel theory \cite{gallager1968information}, which formalized the fundamental limits of temporal-domain communication. Building on this, Franceschetti’s work \cite{Franceschetti2018} extended the classical MIMO paradigm to an electromagnetic framework, establishing a wave theory of information that bridges Maxwell’s equations and Shannon’s capacity. Recent advancements have further explored spatial-domain EIT models, including holographic MIMO (also termed continuous-aperture MIMO) \cite{zhang2023pattern,9475156,9906802,10158997,9650519},  which utilizes a dense array of antennas to create a continuous aperture, enabling the manipulation of electromagnetic waves in three-dimensional space for high-resolution beamforming and increased spectral efficiency \cite{9136592}. Its objective is to fully exploit the propagation characteristics offered by the electromagnetic channel, thereby approaching the fundamental performance limit. In this regard, \cite{Pizzo2020Spatially} and \cite{Wei2022} proposed a Fourier plane-wave series expansion-based channel model for holographic MIMO systems, providing a physically meaningful framework that accurately captures the propagation characteristics of the EM wave. Considering non-ideal propagation environments, \cite{Tengjiao2022} modeled the realistic angular power spectrum based on the von Mises-Fisher (VMF) distributions \cite{4907468} and the 3GPP TR 38.901 channel model \cite{3GPP38901}. Furthermore, the hardware characteristics of antennas, including mutual coupling effects on antenna pattern distortion and efficiency, as well as antenna polarization, are intricately coupled within the holographic MIMO channel model \cite{10500399}.

Second, the theoretical tools of EIT also aim to investigate the physical properties of electromagnetic waves that carry information, with the goal of unveiling new DoFs for information transfer. For instance, near-field communication has been suggested to offer potential benefits in terms of expanding transmission DoFs compared to conventional far-field communication \cite{9903389}. Specifically, near-field communication operates in a region where the distance between the base station (BS) and user equipment (UE) is shorter than the Rayleigh distance \cite{7942128}. When users are in the near-field region, the channel matrix may exhibit higher rank and improved conditioning compared to the far-field scenario~\cite{9903389,7101845,5976389}, provided that the signal processing algorithms are tailored to account for spherical wavefront characteristics\footnote{Note that the improved SNR in near-field scenarios due to shorter distances may confound the analysis of channel matrix conditioning. Our evaluation in Section \ref{sec-NFC} isolates these effects by normalizing transmit power.}. Specifically, near-field communication introduces additional distance-dependent phase variations across the antenna array, enabling the exploitation of spatial DoFs beyond the angular domain \cite{10123941}. However, it is important to note that near-field communication is not an inherently superior option but a condition that may offer advantages in specific scenarios. Additionally, building upon the foundation of EIT, tri-polarized antennas represent a novel approach to further expanding the DoFs in the polarization domain \cite{7270309}. Compared to traditional dual-polarized antennas, tri-polarized antennas allow for the reception and transmission of signals across three orthogonal polarizations, effectively increasing the capacity of the wireless channel \cite{4570210}. It was revealed in \cite{8720148} that tri-polarized antennas can introduce a new orthogonal transmission dimension and improve the eigenvalue distribution of the channel matrix, thereby enhancing channel capacity.

The theoretical framework of EIT has inspired a series of technologies that have been  proven to possess the potential for effectively enhancing spectral efficiency and other communication performance metrics, indicating promising avenues for future research.
Despite significant advancements in the realm of EIT, there remains a notable gap in the practical validation and industrial applicability of theoretical models and analysis.
Specifically, extensive modeling and analysis have been conducted for holographic MIMO, but there is a lack of validation measurements and insights from an industry perspective. 
Besides, in the context of near-field communication, research typically relies on fundamental modeling approaches without considering the statistical characteristics of wireless channels in practical scattering environments and the practical antenna configurations. 
Moreover, existing work on tri-polarized antennas focused on antenna design and hardware characteristics without a comprehensive investigation into its impact on communication systems.

In this paper, we aim to provide a tutorial introduction to the field of Electromagnetic Information Theory (EIT) from the perspective of industrial researchers. Specifically, we try to fill in the above research gaps by investigating the EIT-based MIMO design paradigm, typical applications, validation methods, and future research directions from the perspective of industry.
The main contributions of this paper are summarized as follows:
\begin{itemize}
	\item We propose a novel EIT-based MIMO model that generalizes prior spatially constrained architectures (e.g., holographic MIMO \cite{9475156}, wavenumber-division multiplexing \cite{9906802}) by introducing EM precoding/combining for arbitrary antenna geometries. Our framework leverages dyadic Green’s functions to achieve precision-tunable wave propagation modeling, while unifying hardware imperfections (e.g., mutual coupling) into the channel formulation. Based on the dyadic Green's function and its simplified versions, the proposed model is able to model the EM wave propagation to an arbitrary precision. 
	\item We propose an EM-compliant channel model based on the Fourier plane-wave series, which takes into account the non-isotropic characteristics of the propagation environment, antenna pattern distortion, antenna efficiency, and the polarization. We conduct a channel measurement experiment in an indoor environment and perform channel characteristic analysis based on the real channel datasets.
	\item We extend the 3GPP TR 38.901 channel model by adapting the plane-wave characteristics to the spherical-wave characteristics for near-field communications. In addition, we also incorporates the visible-probability with a power attenuation factor into the near-field channel model to capture the spatial non-stationarity.
	Both theoretical simulation and experimental evaluation with practical array configurations are provided to validate our proposed near-field channel model.	
	\item We consider a practical wireless communication scenario that accounts for the tri-polarized antenna configuration.
	To address the inherent imbalance in receive power among different antenna ports corresponding different polarizations, we propose a joint uplink-downlink channel estimation method.
	Practical antenna patterns, corresponding to the three orthogonal polarizations, are then generated through electromagnetic simulations and integrated into the proposed channel model.
	The efficacy of the proposed channel estimation method is evaluated through system-level simulations within realistic communication scenarios.
\end{itemize}

The rest of this paper is organized as follows. 
The paradigm shifts for future MIMO design based on the EIT framework are discussed in Section II.
Then, the practical channel modeling, validation and analysis for EIT-based typical applications are elaborated in Section III.
The promising research directions and challenges of EIT applications are introduced in Section VI.
Finally, the conclusions and discussions are given in Section V.

\textit{Notation}: Column vectors and matrices are denoted by lowercase and uppercase boldface letters. The imaginary unit is denoted by $j$. The modulus operation is denoted by mod($\cdot$,$\cdot$). $\lfloor \cdot \rfloor$ rounds the argument toward negative infinity. Conjugate transposition is denoted by $(\cdot)^\mathrm{H}$. The cardinality of a set is denoted by $|\cdot|$. $\mathbf{I}$ denotes an identity matrix, while $\mathbf{1}$ and $\mathbf{0}$ denote all-ones and all-zeors matrices. $[\mathbf{A}]_{ij}$ represents the $i,j$ entry of matrix $\mathbf{A}$. The Hadamard and Kronecker products are $\mathbf{A} \odot \mathbf{B}$ and $\mathbf{A} \otimes \mathbf{B}$, respectively. Overlined symbols (e.g., $\bar{\mathbf{x}}$) denote signals at antenna ports, while non-overlined symbols (e.g., $\mathbf{x}$) represent baseband data streams. Noise terms follow the same convention.

\vspace{-10pt}
\section{Paradigm Shifts for Future MIMO}\label{secII}

\begin{figure*}[!t]
	\centering
	\includegraphics[width=6in]{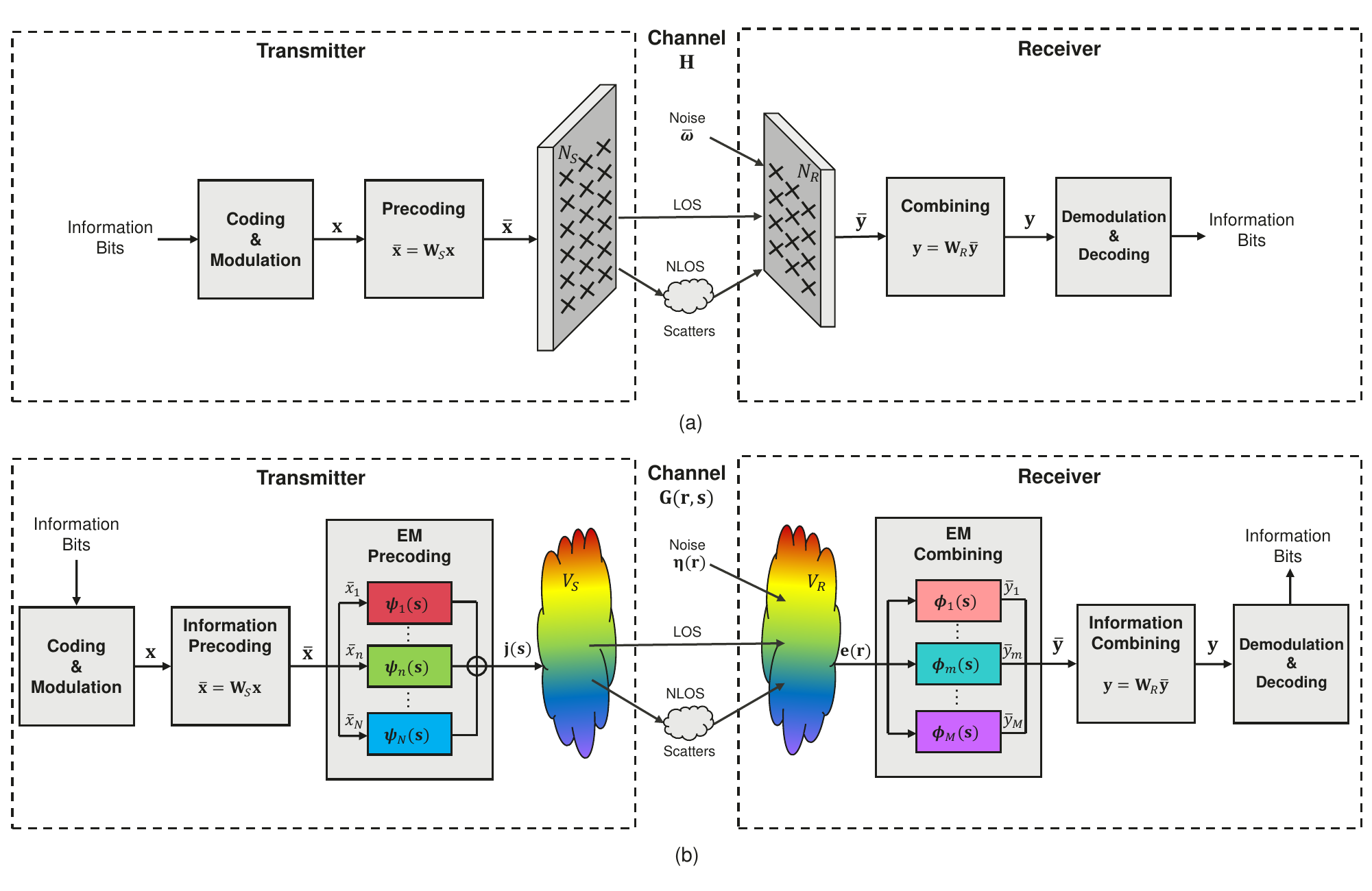}
	\caption{Block diagrams of the conventional MIMO paradigm and the proposed EIT-based MIMO paradigm. a)~Conventional MIMO paradigm; b)~Proposed EIT-based MIMO paradigm.}
	\label{Fig-Model}
\end{figure*}

\subsection{Conventional MIMO Paradigm} \label{Sec-Conv}
In this section, we briefly review the conventional MIMO paradigm, the block diagram of which is shown in Fig.~\ref{Fig-Model}(a). We consider a monochromatic MIMO system with $N_S$ transmit antennas at the transmitter and $N_R$ receive antennas at the receiver. $K$ data streams are transmitted simultaneously~\cite{M-MIMO1}. 

At the transmitter, we denote the data symbols after coding and modulation as $\mathbf{x} \in \mathbb{C}^{K  \times 1}$. Then, a precoding procedure is applied to map $K$ data streams to $N_S$ transmit antennas. Therefore, the transmitted signals can be derived as $\bar{\mathbf{x}} = \mathbf{W}_S \mathbf{x}$, where $\mathbf{W}_S \in \mathbb{C}^{N_S \times K}$ denotes the precoding matrix\footnote{Here, $\bar{\mathbf{x}}$ represents the signal at the antenna ports, while $\mathbf{x}$ is the baseband data stream.}, which can be a digital precoder, analog precoder, or a hybrid digital and analog precoder.

At the receiver, the received signals from $N_R$ antennas are denoted as $\bar{\mathbf{y}} \in \mathbb{C}^{N_R \times 1}$. It can be given as $\bar{\mathbf{y}} = \mathbf{H} \bar{\mathbf{x}} + \bar{\boldsymbol{\omega}}$, where $\mathbf{H} \in \mathbb{C}^{N_R \times N_S}$ denotes the channel matrix and $\bar{\boldsymbol{\omega}} \in \mathbb{C}^{N_R \times 1}$ denotes the noise. Then, a combiner is used to recover $K$ data streams from the received signals. As a result, the overall system can be modeled as 
\begin{equation} \label{Equ-Con}
	\mathbf{y} = \mathbf{W}_R \mathbf{H} \mathbf{W}_S \mathbf{x} + \boldsymbol{\omega},
\end{equation}
where $\mathbf{W}_R \in \mathbb{C}^{K \times N_R}$ is the combining matrix, and $\boldsymbol{\omega} = \mathbf{W}_R \bar{\boldsymbol{\omega}}$ is the equivalent noise. Finally, information bits are extracted from $\mathbf{y}$ through demodulation and decoding.

\subsection{EIT-based MIMO Paradigm}
In the above section, the conventional MIMO paradigm is briefly reviewed, which is a baseband model and spatially-discrete in nature. This model is simple and has directed us in the research of the MIMO system. However, when we are faced with the next generation MIMO system, this model becomes ineffective. For example, it cannot model the holographic MIMO system with spatially-continuous aperture. And the polarization characteristics of the channel and antennas are not modeled. The near-far-field effect of the EM ware are not considered as well.

In this section, we try to bridge this gap by proposing an EIT-based MIMO paradigm, the block diagram of which is shown in Fig.~\ref{Fig-Model}(b). By introducing the concepts of \textit{EM precoding} and \textit{EM combining}, this model provides a general framework for analyzing the performance of the next generation MIMO systems. Several fundamental issues are discussed in this sub-section, including the model of the transmitter, the receiver, and the channel.

\subsubsection{\textbf{Transmitter}}
Generally speaking, in a MIMO system, the information is conveyed by the EM wave emitted from an antenna array at the transmitter. The antenna could be a linear, planar, or a volume array. The length, area, or volume of the aperature at the antenna array is denoted as $V_S$.

We denote the data symbols after coding and modulation as $\mathbf{x} \in \mathbb{C}^{K  \times 1}$. Generally, the data symbols are mapped to different antenna ports. We assume there are $N$ antenna ports at the transmitter. The signals after precoding is given by 
\begin{equation}
	\bar{\mathbf{x}} = \mathbf{W}_S \mathbf{x},
\end{equation}
where $\mathbf{W}_S$ is the precoding matrix.

Actually, in the EM level, each port will induce a current at the transmit antenna aperture. We denote the vector current density function of the $n$-th port as $\boldsymbol{\psi}_n(\mathbf{s}) : \mathbb{R}^3 \rightarrow \mathbb{R}^3$, which is a continuous vector field defined on the 3D spatial domain $V_S$. Here $\mathbf{s} \in \mathbb{R}^{3}$ denotes the coordinate of the transmit aperture. $\boldsymbol{\psi}_n(\mathbf{s})$ is defined as the \textit{EM precoding} in the model, which maps the signals at each port to the vector-valued current density on the transmit aperture. Therefore, the overall current density vector field of the transmit aperture can be derived as:
\begin{equation}
	\mathbf{j}(\mathbf{s}) = \sum_{n=1}^{N} \bar{x}_n \boldsymbol{\psi}_n(\mathbf{s}),
\end{equation}
where $\bar{x}_n$ is the $n$-th element of $\bar{\mathbf{x}}$.



\subsubsection{\textbf{Channel}}
According to the EM theory, for an arbitrary current density function along a radiating surface, the electric field strength $\mathbf{e}(\mathbf{r})$ at the position $\mathbf{r} \in \mathbb{R}^{3 \times 1}$ can be given by
\begin{equation}
	\mathbf{e}(\mathbf{r}) = -j\omega\mu \int_{V_S}\mathbf{\bar G}(\mathbf{r},\mathbf{s})\mathbf{j}(\mathbf{s}) ~\mathrm{d}\mathbf{s} + \boldsymbol{\eta}(\mathbf{r}),
\end{equation}
where $\mathbf{\bar G}(\mathbf{r},\mathbf{s})$ is the dyadic Green's function, $\omega$ is the angular frequency of the EM wave, $\mu$ is the conductivity, respectively. 
$\boldsymbol{\eta}(\mathbf{r}) : \mathbb{R}^3 \rightarrow \mathbb{R}^3$ represents the equivalent noise vector at the receiver, accounting for aggregated thermal noise from active components in the receiver chain (e.g., LNA, mixer, ADC), which is independent of the spatial EM sampling process.

Therefore, when taking the conveyed information at the transmitter into consideration, the electric field strength can be further given by
\begin{equation}
	\begin{split}
		\mathbf{e}(\mathbf{r}) 
	&= -j\omega\mu \int_{V_S}\mathbf{\bar G}(\mathbf{r},\mathbf{s}) \sum_{n=1}^{N} \bar{x}_n \boldsymbol{\psi}_n(\mathbf{s}) ~\mathrm{d}\mathbf{s} + \boldsymbol{\eta}(\mathbf{r}) \\
	&= -j\omega\mu \sum_{n=1}^{N} \bar{x}_n \int_{V_S}\mathbf{\bar G}(\mathbf{r},\mathbf{s})  \boldsymbol{\psi}_n(\mathbf{s}) ~\mathrm{d}\mathbf{s} + \boldsymbol{\eta}(\mathbf{r}).
	\end{split}
\end{equation}

Actually, the calculation of the dyadic Green's function in a real propagation environment is very complicated. However, in a free-space environment, 
the dyadic Green's function can be rigorously defined through the Hessian operator. For a scalar Green's function $G_0: \mathbb{R}^3 \to \mathbb{R}$, we define the Hessian operator $\nabla_{\mathbf{r}}\nabla_{\mathbf{r}}^\top$ acting on the observation coordinate $\mathbf{r}$, where $\nabla_{\mathbf{r}}$ denotes the gradient with respect to $\mathbf{r} \in \mathbb{R}^3$. This leads to:
\begin{equation}
	\mathbf{\bar G}_0 (\mathbf{r},\mathbf{s}) = \left[ \mathbf{{I}} + \frac{\nabla_{\mathbf{r}}\nabla_{\mathbf{r}}^\top}{k_0^2} \right] G_0(\|\mathbf{r}-\mathbf{s}\|)
\end{equation}
where $\nabla_{\mathbf{r}}\nabla_{\mathbf{r}}^\top$ denotes the Hessian matrix operator with respect to $\mathbf{r}$, and $G_0$ is understood as a function of the relative position $\mathbf{p} = \mathbf{r} - \mathbf{s}$,  $k_0 = 2\pi/\lambda$ is the wavenumber of the EM wave, $G_0(\mathbf{p})$ is the scalar Green's function, which can be further given by
\begin{equation} \label{equ-Green-1}
	G_0(\mathbf{p}) = \frac{e^{-j k_0 |\mathbf{p}|}}{4 \pi |\mathbf{p}|}.
\end{equation}

\subsubsection{\textbf{Receiver}} At the receiver, a receiving surface of arbitrary shape is used to collect the EM wave in the environment. The volume of the surface is denoted as $V_R$.
We assume there are $M$ ports on the receiving surface. For the $m$-th port, the received symbol can be seen as a spatial sampling of the electric field strength on the surface. We denote the spatial sampling function as $\boldsymbol{\phi}_m(\mathbf{r})$, which is called \textit{EM combining} at the receiver.

Therefore, the received symbol at the $m$-th port can be derived as
\begin{equation}
	\bar{y}_m = \int_{V_R} \boldsymbol{\phi}^\mathrm{H}_m(\mathbf{r}) \mathbf{e}(\mathbf{r}) ~\mathrm{d}\mathbf{r},
\end{equation}
where $\bar{y}_m$ is the antenna-port-level signal. After combining with $\mathbf{W}_R$, the baseband data stream $\mathbf{y}$ is obtained.
When the transmitted symbols are taken into consideration, the received symbols can be further given by
\begin{equation}
	\begin{split}
		\bar{y}_m 
		=& -j\omega\mu \sum_{n=1}^{N} \bar{x}_n \int_{V_R} \int_{V_S} \boldsymbol{\phi}^\mathrm{H}_m(\mathbf{r}) \mathbf{\bar G}(\mathbf{r},\mathbf{s}) \boldsymbol{\psi}_n(\mathbf{s}) ~\mathrm{d}\mathbf{s}\mathrm{d}\mathbf{r} \\
		&+ \int_{V_R} \boldsymbol{\phi}^\mathrm{H}_m(\mathbf{r}) \boldsymbol{\eta}(\mathbf{r}) ~\mathrm{d}\mathbf{r}.
	\end{split}
\end{equation}


Thus, we define the overall channel from the $n$-th port at the transmitter and the $m$-th port at the receiver as $g_{m,n}$, which can be given by
\begin{equation}
	g_{m,n} \triangleq -j\omega\mu \int_{V_R} \int_{V_S} \boldsymbol{\phi}^\mathrm{H}_m(\mathbf{r}) \mathbf{\bar G}(\mathbf{r},\mathbf{s}) \boldsymbol{\psi}_n(\mathbf{s})~\mathrm{d}\mathbf{s}\mathrm{d}\mathbf{r},
\end{equation}
and the noise is defined as
\begin{equation}
	\bar{\omega}_{m} \triangleq \int_{V_R} \boldsymbol{\phi}^\mathrm{H}_m(\mathbf{r}) \boldsymbol{\eta}(\mathbf{r}) ~\mathrm{d}\mathbf{r},
\end{equation}
we can get the final model as
\begin{equation}
	\bar{y}_m = \sum_{n=1}^N g_{m,n} \bar{x}_n + \bar{\omega}_m.
\end{equation}
When written in a matrix version, the system model can be further given as
\begin{equation}
	\bar{\mathbf{y}} = \mathbf{G}\bar{\mathbf{x}} + \bar{\boldsymbol{\omega}},
\end{equation}
where $\mathbf{G} \in \mathbb{R}^{M \times N}$ is a matrix whose elements are $g_{m,n}$. When the mapping matrix between the symbols and ports at the transmitter $\mathbf{W}_S$ and the receiver $\mathbf{W}_R$ are considered, the overall system model finally becomes
\begin{equation} \label{Equ-EIT}
	\mathbf{y}  = \mathbf{W}_R \mathbf{G} \mathbf{W}_S \mathbf{x} + \boldsymbol{\omega},
\end{equation}
where the equivalent noise is defined as $\boldsymbol{\omega} = \mathbf{W}_R \bar{\boldsymbol{\omega}}$.

\textbf{Remark:} It is worth noting that although equation (\ref{Equ-EIT}) and equation (\ref{Equ-Con}) look similar, actually they have significantly different meanings. In equation (\ref{Equ-Con}), $\mathbf{H}$ is the channel matrix from the $n$-th transmit antenna to the $m$-th receive antenna at specific locations. However, in equation (\ref{Equ-EIT}), $\mathbf{G}$ is the EM-level channel matrix derived from the dyadic Green's function, the EM precoding at the transmitter, and the EM combining at the receiver. This will provide a universal analyzing tool for the next generation MIMO systems, which will be detailed in the following subsections.

\vspace{-10pt}
\subsection{Capacity Calculation}
Based on the system model in~(\ref{Equ-EIT}), we are able to calculate the channel capacity of the EIT-based MIMO system. For a point-to-point communication system, the channel capacity can be derived by
\begin{equation} 
	\begin{split}
	C = &\max_{{Tr(\mathbf{\Xi})}<P} I(\mathbf{x}, \mathbf{y}) \\
	= &\max_{Tr(\mathbf{\Xi})<P} \log_2 \det \left( \mathbf{I} + \frac{1}{\sigma^2} \mathbf{G}\Xi\mathbf{G}^\mathrm{H} \right),
	\end{split}
\end{equation}
where $\Xi = \mathbb{E}(\mathbf{x}\mathbf{x}^\mathrm{H})$ is the covariance matrix of $\mathbf{x}$, and $P$ is the total transmit power.

\vspace{-10pt}
\subsection{Consistency with Conventional MIMO Model}
In this subsection, we will show how the proposed model is consistent with the conventional MIMO model in Section~\ref{Sec-Conv}. In conventional MIMO model, all the signals are modeled in a base-band level without considering the EM-level characteristics. 

Therefore, the conventional MIMO model can be regarded as that each port excites a point source at the transmit surface, which can be modeled as a delta function.
If we denote the location of each transmit antenna as $\mathbf{s}_n \in \mathbb{R}^{3 \times 1}$, 
the EM precoding for the conventional MIMO can be defined as  
\begin{equation}
	\tilde{\boldsymbol{\psi}}_n(\mathbf{s}) \triangleq \boldsymbol{\delta}(\mathbf{s} - \mathbf{s}_n) \cdot \mathbf{p}_n,
\end{equation}
where $\mathbf{p}_n \in \mathbb{R}^3$ denotes the polarization vector of the $n$-th antenna.
Similarly, at the receiver, the EM combining function can be defined as 
\begin{equation}
	\tilde{\boldsymbol{\phi}}_m(\mathbf{r}) \triangleq \boldsymbol{\delta}(\mathbf{r} - \mathbf{r}_m),
\end{equation}
where $\mathbf{r}_m \in \mathbb{R}^{3 \times 1}$ denotes the locations of the receive antennas.

As a result, the channel from the $n$-th transmit port to the $m$-th receive port can be derived as
\begin{equation}
	\begin{split}
		\tilde{g}_{m,n} 
		&= -j\omega\mu \int_{V_R} \int_{V_S} \boldsymbol{\delta}(\mathbf{r} - \mathbf{r}_m)^\mathrm{H} \mathbf{G}(\mathbf{r},\mathbf{s}) \boldsymbol{\delta}(\mathbf{s} - \mathbf{s}_n)~\mathrm{d}\mathbf{s}\mathrm{d}\mathbf{r} \\
		&= -j\omega\mu \mathbf{G}(\mathbf{r}_m, \mathbf{s}_n),
	\end{split}
\end{equation}
which means that the channel for conventional MIMO is the Green's function from the transmit antenna at $\mathbf{s}_n$ to the receive antenna at $\mathbf{r}_m$. This result is also consistent with the EM wave propagation.

Thus the equivalent noise will become
\begin{equation}
	\tilde{\omega}_{m} = \int_{V_R} \boldsymbol{\delta}(\mathbf{r} - \mathbf{r}_m)^\mathrm{H} \boldsymbol{\eta}(\mathbf{r}) ~\mathrm{d}\mathbf{r} = \boldsymbol{\eta}(\mathbf{r}_m),
\end{equation}
which means that the equivalent noise is the noise at each receive antenna located at $\mathbf{r}_m$.

Finally, the EIT-based model will be degenerated into
\begin{equation}
	\mathbf{y} = \mathbf{W}_R \tilde{\mathbf{G}} \mathbf{W}_S \mathbf{x} + \hat{\boldsymbol{\omega}},
\end{equation}
which is consistent with the model in equation~(\ref{Equ-Con}) in Section~\ref{Sec-Conv}.

\subsection{Universal Model for the Next Generation MIMO}

\begin{figure*}[!t]
	\centering
	\includegraphics[width=6in]{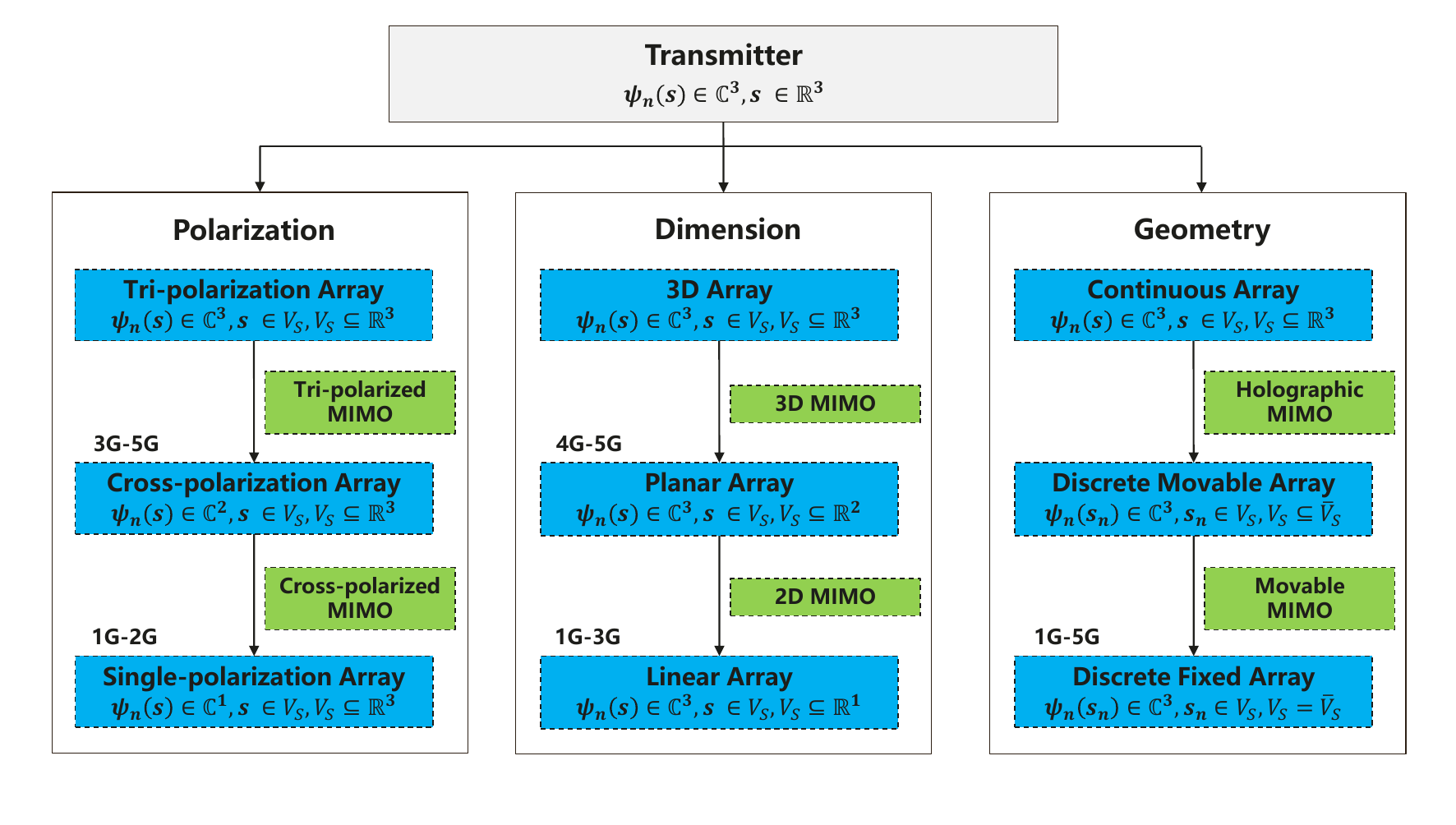}
	\caption{The proposed EIT-based model is able to account for arbitrary shapes of transmit antennas at the base station.}
	\label{Fig-Transmitter}
\end{figure*}

In this subsection, we will show that how the proposed EIT-based model can be seen as a universal model for the next generation MIMO systems. Specifically, the proposed model is able to model arbitrary shapes of the transmit and receive antennas, and to model the EM wave propagation to an arbitrary precision.

\subsubsection{\textbf{Arbitrary Shapes of Antennas}} Based on the proposed EM precoding and EM combining functions, the proposed model is able to model arbitrary shapes of antenna arrays. This is because any antenna array is composed of a set of points in the environment. Specifically, for any practical transmit antenna array with finite volume, we denote the set of the points within the antenna array as $V_S$, which means $\mathbf{s} \in V_S$. At the receiver, the situations are similar, the receive antenna array can be seen as a combinaitons of points defined as $V_R$, and we have $\mathbf{r} \in V_R$.

From 1G to 5G and the future 5G-Advanced MIMO systems, many kinds of antenna arrays are used or under investigation. Alomost any one of the them can be seen as a special case of the proposed model. Specifically, for the transmit antenna array, there exist three domains, including the polarizaiton, the dimension of the array, and the geometry of antenna elements. A block diagram is shown in Fig.~\ref{Fig-Transmitter}.
\begin{itemize}
	\item \textbf{Polarization:} Actually, there exist three orthogonal polarizations in total. In the proposed model, each polarizaiton is expressed by one element of $\boldsymbol{\psi}_n(\mathbf{s})$. Therefore, for the single-polarizaiton array, we have $\boldsymbol{\psi}_n(\mathbf{s}) \in \mathbb{C}^1$. And for the cross-polarizaiton array and the tri-polarizaiton array, we have $\boldsymbol{\psi}_n(\mathbf{s}) \in \mathbb{C}^2$ and $\boldsymbol{\psi}_n(\mathbf{s}) \in \mathbb{C}^3$, respectively. In the wireless industry, from 1G to 2G, single-polarizaiton arrays are always used. And from 3G to 5G, cross-polarizaiton arrays are the main architectures. In the future 5G-Advanced communications, tri-polarizaiton array may be one of the candidate architectures~\cite{4570210,8720148}. We have done some research on the channel acquisition of this array, which will be detailed in Section~\ref{sec-tri}.
	\item \textbf{Dimension:} The dimension of the antenna array is also one of the domains in the design of the transmitter. From 1G to 3G, linear arrays are the main architectures in wireless communications, which can be modeled by $V_S \subseteq \mathbb{R}^1$. From 4G to 5G, the planar arrays are utilized to realize beamforming in both the horizontal and vertical domains, in order to futher improve the spatial multiplexing of MIMO. In the proposed model, this can be expressed by $V_S \subseteq \mathbb{R}^2$. In the furute, maybe three-dimensional arrays will be used. Recent research has shown the benefits of the three-dimensional antenna arrays~\cite{yuan2024breaking}.
	\item \textbf{Geometry:} The geometry of the antenna element is another domain. From 1G to 5G wireless communication systems, we are stilling using the discrete arrays with antenna elements fixed at specific positions. In the proposed model, this corresponds to $\mathbf{s}_n \in V_S, V_S = \bar{V}_S$, which means that the antenna elements are placed at a fixed point set $\bar{V}_S$. For the future 5G-Advanced communications, the movable arrays~\cite{zhu2023movable} and the continuous-aperture arrays~\cite{zhang2023pattern} may be promising candidate architectures. The movable array can be modeled as $\mathbf{s}_n \in V_S, V_S \subseteq \bar{V}_S$, and the continuous arrays can be modeled as $\mathbf{s} \in V_S$. In practice, the continuous antenna array is not easy to frabicate. Densely spaced array may be one important step forward to the target of a complete continuous array. In Section~\ref{sec-holo}, we will also detail our contributions in this field.
\end{itemize}

\vspace{-10pt}
\subsubsection{\textbf{Arbitrary Precisions of Channel Characterization}}
The proposed EIT-based model is also able to describe the propagation channel to arbitrary precisions. The exact EM channel can be derived by the Green's function method based on the Maxwell's functions and boundary conditions in a specific environment. This can be done through full wave simulation by EM calculation softwares. Or we can use the raytracing method to approximate the results to decrease the complexity. These deterministic methods can calculate the channel exactly. However, the complexity is very high and not easy to emplement in a practical communication system.

In the practical communication systems, we always use semi-deterministic method to model the channel in order to decrease the complexity. From 1G to 5G, the industrial channel model is always based on the far-field assumptions, such as the latest 3GPP TR 38.901 channel model~\cite{3GPP38901}. However, for the 5G-Advanced communication system, this assumption may not hold because higher frequency bands and larger antenna arrays are used. In the following, we will show how the proposed model is able to model the channel for the far-field users and also the near-field users in a free-space environment.

Let $\mathbf{R} = \mathbf{r} - \mathbf{s} \in \mathbb{R}^3$ denote the relative position vector with $R = \|\mathbf{R}\|$, and the free-space Green's function can be derived by
\begin{equation}
	\mathbf{\bar G}_0 (\mathbf{r},\mathbf{s}) = \left[ \mathbf{{I}} + \frac{\nabla_{\mathbf{R}}\nabla_{\mathbf{R}}^\top}{k_0^2} \right] \frac{e^{-j k_0 R}}{4 \pi R},
\end{equation}
which can be further given by
\begin{equation}
\begin{split}
	\mathbf{\bar G}_0 (\mathbf{r},\mathbf{s}) =
	 &\frac{e^{j k_0 R}}{4 \pi R}  \mathbf{{I}} \left(- \frac{4 \pi}{3 k_0^2} \delta(\mathbf{R}) + 1 + \frac{j}{kR} - \frac{1}{k_0^2 R^2}   \right) \\
	 &+ \frac{e^{j k_0 R}}{4 \pi R} \frac{\mathbf{R}\mathbf{R}^\top}{R^2} \left( -1 - \frac{3j}{k_0 R} + \frac{3}{k_0^2 R^2} \right).
\end{split}
\end{equation}

Therefore, for $\mathbf{R} \neq \mathbf{0}$, the free-space Green's function can be decomposed into three terms
\begin{align}
	\mathbf{G}_{\mathrm{INF}} &= \frac{e^{-jk_0 R}}{4\pi k_0^2 R^3} \left(-\mathbf{I} + 3\frac{\mathbf{R}\mathbf{R}^\top}{R^2}\right), \\
	\mathbf{G}_{\mathrm{RNF}} &= \frac{j e^{-jk_0 R}}{4\pi k_0 R^2} \left(\mathbf{I} - 3\frac{\mathbf{R}\mathbf{R}^\top}{R^2}\right), \\
	\mathbf{G}_{\mathrm{FF}} &= \frac{e^{-jk_0 R}}{4\pi R} \left(\mathbf{I} - \frac{\mathbf{R}\mathbf{R}^\top}{R^2}\right).
\end{align}
These three terms help to divide the EM field into three regions, including the reactive near field, the radiating near field, and the far field. According to~\cite{7942128}, the boundary of the reactive near-field region can be given by $0.62\sqrt{D^3/\lambda}$, and the boundary of the radiating near-field region is the Rayleigh distance $2D^2/\lambda$.

\begin{figure}[!t]
	\centering
	\includegraphics[width=3.5 in]{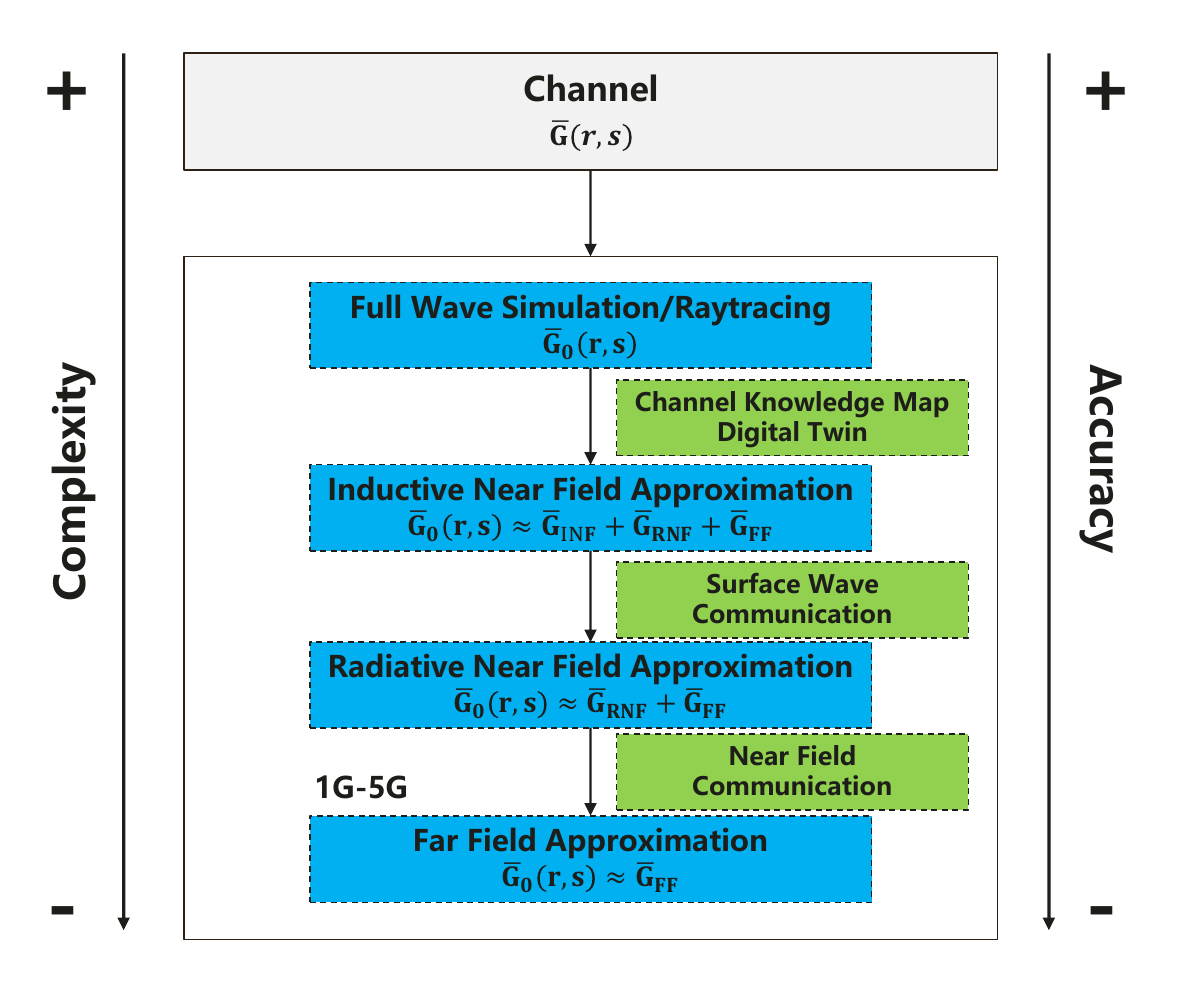}
	\caption{The proposed EIT-based model is able to model arbitrary precisions of the channel characterization.}
	\label{Fig-Channel}
\end{figure}

The whole picture can be seen in Fig.~\ref{Fig-Channel}, which shows that the proposed model is able to describe the channel to different levels of precisions. In the future, the near-field communication~\cite{9903389}, surface wave communication~\cite{wong2020vision}, and channel knowledge map~\cite{zeng2021toward} may be key techniques to better utilize the channel characteristics to improve the system performance. In Section~\ref{sec-NFC}, we will also explain our recent research on the near-field communications.

\section{Applications, Modelings, and Numerical Results}

In this section, we explore typical paradigms based on the EIT theory, specifically focusing on densely spaced MIMO, near-field communications, and tri-polarized MIMO. 
Considering the practical scattering environment and antenna configurations from the industrial perspective, we introduce the channel modeling and analyze the channel characteristics of mentioned applications.
Numerical results are provided based on the theoretical simulation and measured channel datasets to evaluate the performance of mentioned techniques.

\vspace{-10pt}
\subsection{Densely Spaced MIMO} \label{sec-holo}

In this subsection, we study an EM-compliant channel model for densely spaced array based on the principle of EIT by taking into account the practical propagation environment and mutual coupling effects \footnote{In the extended model, as compared to \cite{10500399,wan2023mutualinformationelectromagneticinformation,4636839}, not only the non-isotropic characteristics of the propagation environment, the antenna pattern distortion, the antenna efficiency, but also the polarization of the antennas and the propagation environment can be modeled.}. 
Based on the proposed channel model, the real-world channel for densely spaced MIMO systems is measured in an indoor environment.
The channel capacity of a densely spaced array systems is evaluated according to both the theoretical model and measurement results.

\subsubsection{\textbf{Channel Model}}

The proposed channel model in this section can be viewed as a specialized implementation of the universal EIT-based framework in Section \ref{secII}.
We consider $N_\mathrm{R}$ and $N_\mathrm{S}$ antenna elements on the planar antenna arrays with the size of $L_\mathrm{R}^x\times L_\mathrm{R}^y$ ($L_\mathrm{S}^x\times L_\mathrm{S}^y$) at the transmitter (Tx) and receiver (Rx), respectively. 
The distances between adjacent antenna elements are denoted by $\{ \Delta_\mathrm{R}^x, \Delta_\mathrm{R}^y \}$ and $\{ \Delta_\mathrm{S}^x, \Delta_\mathrm{S}^y \}$ for antenna arrays at the receiver and transmitter, respectively. 
The continuous aperture $V_S$ and $V_R$ are constrained to planar arrays with discrete elements at positions $\{\mathbf{s}_p\}$ and $\{\mathbf{r}_q\}$. As such, the locations of each antenna elements are denoted by $\mathbf{r}_q \triangleq (r_q^x, r_q^y, r_q^z)$ with $q = 1,2,\cdots,N_\mathrm{R}$ and $\mathbf{s}_p \triangleq (s_p^x, s_p^y, s_p^z))$ with $p = 1,2,\cdots,N_\mathrm{S}$, respectively. 
Note that under limited antenna aperture conditions, as the number of antenna elements approaches infinity, the densely spaced array transitions into a holographic MIMO with a continuous aperture.

First, the basic wavenumber-domain channel $\mathbf{H}_a \in \mathbb{C}^{n_\mathrm{R} \times n_\mathrm{S}}$ can be modeled by a set of random Fourier coefficients, $[\mathbf{H}_\mathrm{a}]_{\beta, \alpha}$ follows the complex Gaussian distribution $\mathcal{CN}(\mu_{\beta, \alpha}, \sigma^2_{\beta, \alpha})$. Here, $\mu_{\beta, \alpha}$ represents deterministic contributions (e.g., LoS/specular reflections), while $\sigma^2_{\beta, \alpha}$ models diffuse scattering. For simplicity, we focus on $\mu_{\beta, \alpha} = 0$ in scattering-dominated environments, but $\mu_{\beta, \alpha}$ can be explicitly incorporated using ray-tracing or geometric optics in LoS-dominant cases.
\begin{equation}
	\mathcal{E}_\mathrm{R} = \left\{ (l^x, l^y) \in \mathbb{Z}^2: \left( \frac{l^x \lambda}{L_\mathrm{R}^x} \right)^2 + \left( \frac{l^y \lambda}{L_\mathrm{R}^y} \right)^2 \leq 1 \right\},
\end{equation}
\begin{equation}
	\mathcal{E}_\mathrm{S} = \left\{ (m^x, m^y) \in \mathbb{Z}^2: \left( \frac{m^x \lambda}{L_\mathrm{S}^x} \right)^2 + \left( \frac{m^y \lambda}{L_\mathrm{S}^y} \right)^2 \leq 1 \right\}.
\end{equation}
The variance $\sigma^2_{\beta, \alpha}$ is given by
\begin{equation}
	\begin{split}
		\sigma^2_{\beta, \alpha} = & \int \!\!\! \int_{\Omega_\mathrm{R}(l^x_{\beta}, l^y_{\beta})} \!\!\! A^2(\theta_\mathrm{R}, \phi_\mathrm{R}) \sin \theta_\mathrm{R} \mathrm{d}\theta_\mathrm{R} \mathrm{d}\phi_\mathrm{R} \\
		& \int \!\!\! \int_{\Omega_\mathrm{S}(m^x_\alpha, m^y_\alpha)} \!\!\! A^2(\theta_\mathrm{S}, \phi_\mathrm{S}) \sin \theta_\mathrm{S} \mathrm{d}\theta_\mathrm{S} \mathrm{d}\phi_\mathrm{S},
	\end{split}
\end{equation}
where $A^2(\theta_\mathrm{R}, \phi_\mathrm{R})$ and $A^2(\theta_\mathrm{S}, \phi_\mathrm{S})$ denote the angular power spectrum \cite{6542746} at the receiver and the transmitter, respectively. The angular power spectrum can be further modeled by a mixture of VMF distributions~\cite{4907468} as
\begin{equation}
	A^2_\mathrm{R} (\theta_\mathrm{R}, \phi_\mathrm{R}) = \sum_{i=1}^{N_c} w_{\mathrm{R},i} p_{\mathrm{R},i}(\theta_\mathrm{R}, \phi_\mathrm{R}),
\end{equation}
where $N_c$ denotes the number of clusters of the scatters in propagation environment. $w_{\mathrm{R},i}$ denotes the normalization factor with $\sum_i^{N_c} w_{\mathrm{R},i} =  1$. $p_{\mathrm{R},i}(\theta_\mathrm{R}, \phi_\mathrm{R})$ denotes the probability function of the VMF distribution, which can be further expressed as
\begin{equation}
	\begin{split}
		p_{\mathrm{R},i}&(\theta_\mathrm{R}, \phi_\mathrm{R}) = \frac{\alpha_{\mathrm{R},i}}{4 \pi \mathrm{sinh} (\alpha_{\mathrm{R},i})} \\
		& e^{\left( {\alpha_{\mathrm{R},i}(\sin \theta_\mathrm{R} \sin \bar{\theta}_{\mathrm{R},i} \cos(\phi_\mathrm{R} - \bar{\phi}_{\mathrm{R},i}) + \cos \theta_\mathrm{R} \cos \bar{\theta}_{\mathrm{R},i})}\right)} ,
	\end{split}
\end{equation}
where $\{\bar{\phi}_{\mathrm{R},i}, \bar{\theta}_{\mathrm{R},i}\}$ denotes the elevation and azimuth angles of the $i$-th cluster at the receiver, and $\alpha_{\mathrm{S},i}$ denotes the concentration parameters for the $i$-th cluster. These angles can be derived from the 3GPP TR 38.901 channel model~\cite{3GPP38901} according to \cite{Tengjiao2022}. The angular power spectrum at the transmitter can be modeled similarly and omitted for brevity.

Then, considering the polarization characteristics of propagation environment, the co-polarization wavenumber-domain channels, $\mathbf{H}_\mathrm{a}^{\theta\theta}$ and $\mathbf{H}_\mathrm{a}^{\phi\phi}$, can be modeled by involving a random phase shift into the basic wavenumber-domain channels, $\mathbf{H}_a$, as
\begin{equation} 
	[\mathbf{H}_{\mathrm{a}}^{\theta\phi}]_{\beta, \alpha} = [\mathbf{H}_{\mathrm{a}}]_{\beta, \alpha} \, e^{j\Phi^{\theta\phi}_{\beta, \alpha}} \sqrt{\kappa^{-1}_{\beta, \alpha}},
\end{equation}
\begin{equation} 
	[\mathbf{H}_{\mathrm{a}}^{\phi\theta}]_{\beta, \alpha} = [\mathbf{H}_{\mathrm{a}}]_{\beta, \alpha} \, e^{j\Phi^{\phi\theta}_{\beta, \alpha}} \sqrt{\kappa^{-1}_{\beta, \alpha}},
\end{equation}
where $\Phi^{\theta\theta}_{\beta, \alpha}$, $\Phi^{\phi\phi}_{\beta, \alpha}$, $\Phi^{\phi\theta}_{\beta, \alpha}$, and $\Phi^{\theta\phi}_{\beta, \alpha}$ are random phase shifts following the uniform distribution within $[-\pi, \pi]$. $\kappa_{\beta, \alpha}$ denotes the cross polarization power ratios of the propagation environment, following the log-normal distribution $\kappa_{\beta, \alpha} = 10^{X_{\beta, \alpha} / 10}$ with $X_{\beta, \alpha} \sim \mathcal{N}(\mu_{\mathrm{XPR}}, \sigma^2_\mathrm{XPR})$.

Next, the polarization distortion of antennas can be modeled by embedding the element directivity patterns into the Fourier harmonics.
The Fourier harmonics implement spatial sampling constrained by array apertures $L_\mathrm{R}^x \times L_\mathrm{R}^y$ and $L_\mathrm{S}^x \times L_\mathrm{S}^y$, translating the continuous EM integration to discrete matrix products.
Specifically, the modified Fourier harmonics with the horizontal and vertical polarization at the receiver can be expressed as
\begin{equation} 
	\begin{split}
		[\boldsymbol{\Psi}_\mathrm{R}^{\theta}]_{q,\beta} =
		& \frac{1}{\sqrt{N_\mathrm{R}}} e^{j\left( \frac{2\pi l^x_\beta}{L_\mathrm{R}^x} r^x_q + \frac{2\pi l^y_\beta}{L_\mathrm{R}^y} r^y_q + \gamma_\mathrm{R}(l^x_\beta,l^y_\beta) r_q^z\right)} \\
		& F_{\mathrm{R},q}^{\theta}\left( \hat{\theta}_\mathrm{R}(l^x_\beta, l^y_\beta) , \hat{\phi}_\mathrm{R}(l^x_\beta, l^y_\beta) \right),
	\end{split}
\end{equation}
and
\begin{equation} 
	\begin{split}
		[\boldsymbol{\Psi}_\mathrm{R}^{\phi}]_{q,\beta} =
		& \frac{1}{\sqrt{N_\mathrm{R}}} e^{j\left( \frac{2\pi l^x_\beta}{L_\mathrm{R}^x} r^x_q + \frac{2\pi l^y_\beta}{L_\mathrm{R}^y} r^y_q + \gamma_\mathrm{R}(l^x_\beta,l^y_\beta) r_q^z\right)} \\
		& F_{\mathrm{R},q}^{\phi}\left( \hat{\theta}_\mathrm{R}(l^x_\beta, l^y_\beta) , \hat{\phi}_\mathrm{R}(l^x_\beta, l^y_\beta) \right),
	\end{split}
\end{equation}
where $\gamma_\mathrm{R}(l^x, l^y) = \sqrt{(\frac{2\pi}{\lambda})^2 - (\frac{2\pi l^x}{L_\mathrm{R}^x})^2 - (\frac{2\pi l^y}{L_\mathrm{R}^y})^2}$ and $\gamma_\mathrm{S}(m^x, m^y) = \sqrt{(\frac{2\pi}{\lambda})^2 - (\frac{2\pi m^x}{L_\mathrm{S}^x})^2 - (\frac{2\pi m^y}{L_\mathrm{S}^y})^2}$.
$F_{\mathrm{R},q}^{\theta} \left( \theta_\mathrm{R}, \phi_\mathrm{R} \right)$ and $F_{\mathrm{R},q}^{\phi} \left( \theta_\mathrm{R}, \phi_\mathrm{R} \right)$ denote the embedded element directivity patterns in the horizontal and the vertical polarization, respectively.
As such, the modified Fourier harmonics at the transmitter, $\boldsymbol{\Psi}_\mathrm{S}^{\phi}$ and $\boldsymbol{\Psi}_\mathrm{S}^{\phi}$, can be expressed similarly with $F_{\mathrm{S},p}^{\theta} \left( \theta_\mathrm{S}, \phi_\mathrm{S} \right)$ and $F_{\mathrm{S},p}^{\phi} \left( \theta_\mathrm{S}, \phi_\mathrm{S} \right)$ denoting the respective embedded element directivity patterns.
Note that the elevation and azimuth angles $\{ \hat{\phi}_\mathrm{R}(l^x, l^y) , \hat{\theta}_\mathrm{R}(l^x, l^y) \}$ and $\{ \hat{\phi}_\mathrm{S}(m^x, m^y) , \hat{\theta}_\mathrm{S}(m^x, m^y) \}$ are derived from $(l^x, l^y)$ and $(m^x, m^y)$ by transformation from the wavenumber domain to the angular domain.

Furthermore, $\boldsymbol{\Gamma}_\mathrm{R} \in \mathbb{R}^{N_\mathrm{R} \times N_\mathrm{R}}$ and $\boldsymbol{\Gamma}_\mathrm{S} \in \mathbb{R}^{N_\mathrm{S} \times N_\mathrm{S}}$ are diagonal matrices representing the efficiency of the antenna element at the receiver and the transmitter, respectively.
Note that the efficiency matrices $\boldsymbol{\Gamma}_\mathrm{R},\boldsymbol{\Gamma}_\mathrm{S}$ and embedded patterns $F^{\theta/\phi}$ model non-ideal antenna characteristics missing in the ideal universal model.

As a result, the full channel matrix $\mathbf{H} \in \mathbb{C}^{N_\mathrm{R} \times N_\mathrm{S}}$ can be expressed as
\begin{equation} 
	\begin{split}
		\mathbf{H} = 
		& \boldsymbol{\Gamma}_\mathrm{R} \boldsymbol{\Psi}_\mathrm{R}^{\theta} {\mathbf{H}}_\mathrm{a}^{\theta\theta} {\boldsymbol{\Psi}_\mathrm{S}^\theta}^\mathrm{H} \boldsymbol{\Gamma}_\mathrm{S} + \boldsymbol{\Gamma}_\mathrm{R} \boldsymbol{\Psi}_\mathrm{R}^{\theta} {\mathbf{H}}_\mathrm{a}^{\theta\phi} {\boldsymbol{\Psi}_\mathrm{S}^\phi}^\mathrm{H} \boldsymbol{\Gamma}_\mathrm{S}\\
		& + \boldsymbol{\Gamma}_\mathrm{R} \boldsymbol{\Psi}_\mathrm{R}^{\phi} {\mathbf{H}}_\mathrm{a}^{\phi\theta} {\boldsymbol{\Psi}_\mathrm{S}^\theta}^\mathrm{H} \boldsymbol{\Gamma}_\mathrm{S} + \boldsymbol{\Gamma}_\mathrm{R} \boldsymbol{\Psi}_\mathrm{R}^{\phi} {\mathbf{H}}_\mathrm{a}^{\phi\phi} {\boldsymbol{\Psi}_\mathrm{S}^\phi}^\mathrm{H} \boldsymbol{\Gamma}_\mathrm{S} \\
		= & \boldsymbol{\Gamma}_\mathrm{R} 
		\begin{bmatrix} \boldsymbol{\Psi}_\mathrm{R}^{\theta}, \boldsymbol{\Psi}_\mathrm{R}^{\phi}  \end{bmatrix}
		\begin{bmatrix} \mathbf{H}_\mathrm{a}^{\theta\theta} & \mathbf{H}_\mathrm{a}^{\theta\phi} \\ \mathbf{H}_\mathrm{a}^{\phi\theta} & \mathbf{H}_\mathrm{a}^{\phi\phi} \end{bmatrix}
		\begin{bmatrix} \boldsymbol{\Psi}_\mathrm{S}^{\theta}, \boldsymbol{\Psi}_\mathrm{S}^{\phi} \end{bmatrix}^{\mathrm{H}} \boldsymbol{\Gamma}_\mathrm{S}.
	\end{split}
\end{equation}
For clarity, we rewrite the channel matrix in a more compact form as follows
\begin{equation} \label{equ-chn-pol-mat}
	\mathbf{H} = \boldsymbol{\Gamma}_\mathrm{R} \boldsymbol{\Psi}_\mathrm{R}^{\mathrm{pol}} {\mathbf{H}}_\mathrm{a}^{\mathrm{pol}} {\boldsymbol{\Psi}_\mathrm{S}^{\mathrm{pol}}}^\mathrm{H} \boldsymbol{\Gamma}_\mathrm{S},
\end{equation}
where $\boldsymbol{\Psi}_\mathrm{R}^{\mathrm{pol}} = [ \boldsymbol{\Psi}_\mathrm{R}^{\theta}, \boldsymbol{\Psi}_\mathrm{R}^{\phi} ]$, $\boldsymbol{\Psi}_\mathrm{S}^{\mathrm{pol}} = [ \boldsymbol{\Psi}_\mathrm{S}^{\theta}, \boldsymbol{\Psi}_\mathrm{S}^{\phi} ]$, and $\mathbf{H}_\mathrm{a}^{\mathrm{pol}} = \begin{bmatrix} \mathbf{H}_\mathrm{a}^{\theta\theta} & \mathbf{H}_\mathrm{a}^{\theta\phi} \\ \mathbf{H}_\mathrm{a}^{\phi\theta} & \mathbf{H}_\mathrm{a}^{\phi\phi} \end{bmatrix}$. 

Consequently, the non-ideal propagation environment and antenna characteristics are all taken into consideration in the proposed channel model.

\subsubsection{\textbf{Simulation setup}}
 
In the simulation part, we consider the central frequency of $f_c = 4.7$~GHz. 
The width and height of the antenna arrays at BS and UE are fixed as $\{ L_{\mathrm{S},x}, L_{\mathrm{S},y}\} = \{4 \lambda, 4 \lambda\}$ and $\{ L_{\mathrm{R},x}, L_{\mathrm{R},y}\} = \{\lambda, \lambda\}$, where the antenna elements are uniformly placed with $\Delta_\mathrm{S} = \Delta_{\mathrm{S},x} = \Delta_{\mathrm{S},y}$ and $\Delta_\mathrm{R} = \Delta_{\mathrm{R},x} = \Delta_{\mathrm{R},y}$.
Based on the channel model in (\ref{equ-chn-pol-mat}), the angular power spectrum in the non-isotropic scattering environment is simulated according to the CDL-B channel in 3GPP TR 39.901 with $N_c = 23$ clusters, and that in the isotropic scattering environment is simulated by setting $N_c = 1$ and $\alpha_{\mathrm{S},i} = \alpha_{\mathrm{R},i} = 0$.
The antenna pattern distortion is generated from the electromagnetic computing softwares with each element modeled as a dipole antenna.
For the sake of statistical accuracy and robustness, the ergodic capacity is analyzed through Monte Carlo simulations with 1000 channel realizations.

\subsubsection{\textbf{Measurement Setup}}

\begin{figure}[!t]\centering
	\includegraphics[width=0.4\textwidth]{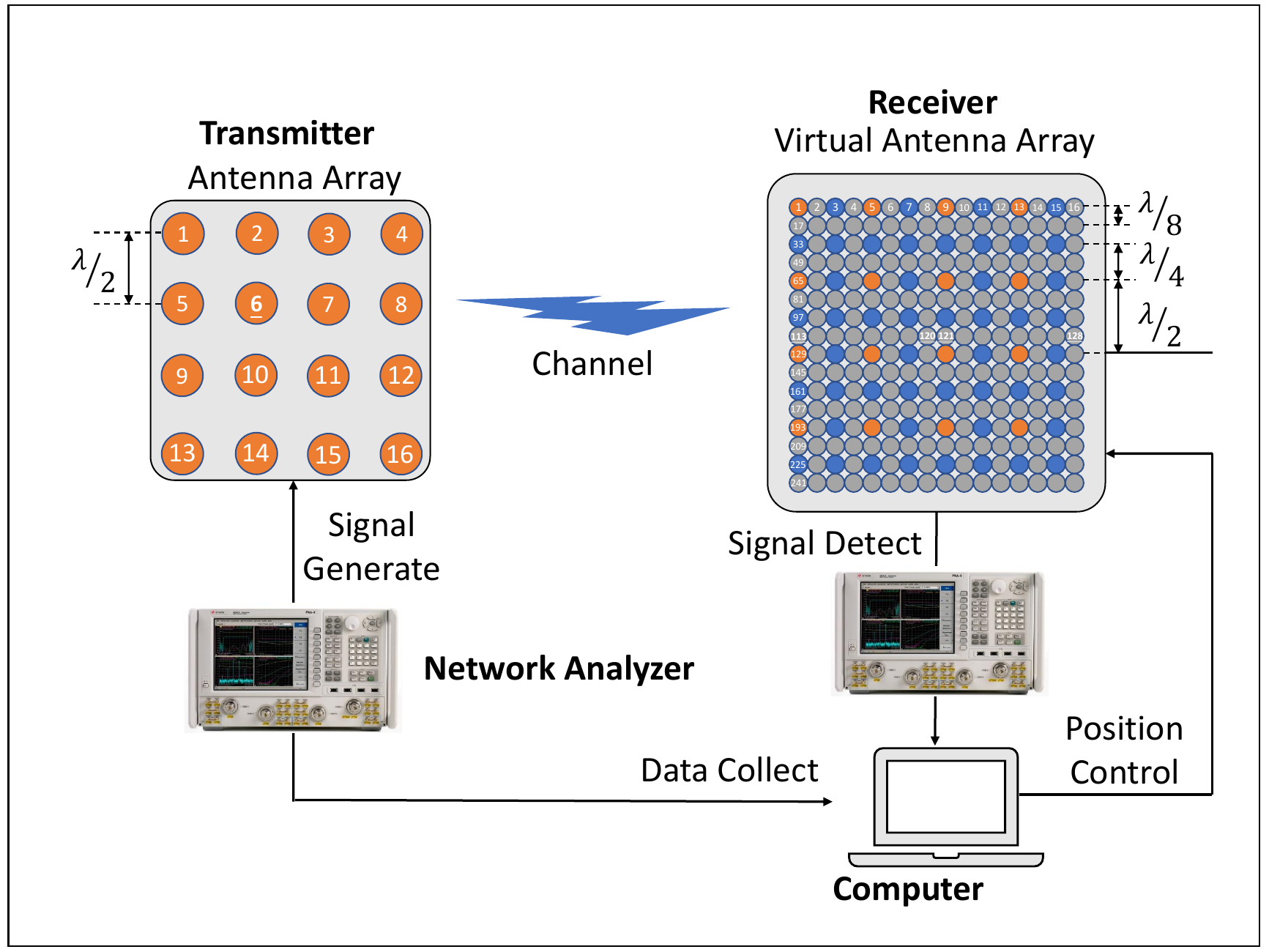}
	\caption{Schematic diagram of the measurement equipment.}
	\label{tx_rx}
\end{figure}

In order to evaluate the performance of densely spaced MIMO systems under the realistic scenarios, an experiment is conducted to measure the real-world channel. 
The schematic diagram of the measurement equipment is shown in Fig.~\ref{tx_rx}.
 
In the experimental setup, the center frequency is set to $f_c = 4.7$ GHz, and the bandwidth spans from $4.6$ GHz to $4.8$ GHz with a total of $1023$ frequency samples. 
A virtual antenna array is employed to emulate the dense antenna array. A discone antenna with an omnidirectional pattern is utilized and controlled by the electrical mechanism. 
Through computer-controlled programming, the antenna is moved to various locations to construct a virtual dense array with customizable element spacings. During the experiment, the virtual antenna array is positioned at the receiver side to implement a densely spaced MIMO array with spacings of $\Delta_\mathrm{R}^x = \Delta_\mathrm{R}^y \in \{\lambda /8, \lambda /4, \lambda /2\}$. At the transmitter, a standard antenna array comprising $N_{\mathrm{S}} = 16$ antennas with spacings of $\Delta_\mathrm{S}^x = \Delta_\mathrm{S}^y = \lambda /2$ is deployed. The antenna elements are patch antennas with a half-power beamwidth of 70$^\circ$.
The transmitters were placed 5.65 meters (m) apart from the receivers. A medal object is placed between them to form an indoor multipath environment.
The signal to noise ratio (SNR) is set to 0 dB. The equal power allocation strategy ensures uniform transmit power across all antennas, isolating the impact of antenna spacing on capacity.
A calibrated network analyzer is used for channel measurement, and the measurement results are collected by a computer. 
Note that the virtual antenna array cannot capture the mutual coupling effects precisely, thus we estimate the efficiency of the antenna element to reflect the effect of mutual coupling \cite{Hannan1964} by
\begin{equation} \label{equ-Hannan}
	\eta_{\mathrm{R},q} \approx \frac{\pi \Delta_\mathrm{R}^x \Delta_\mathrm{R}^y}{\lambda^2}.
\end{equation}
The antenna efficiency $\eta_{\mathrm{R},q}$ in (\ref{equ-Hannan}) is derived from the effective aperture theory, where the efficiency of a densely spaced element is approximated as proportional to its physical area relative to $\lambda^2$. While this simplification does not account for mutual coupling or pattern distortion, it provides a tractable metric for evaluating efficiency loss in dense arrays\footnote{While the current work establishes the foundational coupling-capacity relationship, a comprehensive verification using physical arrays with tunable spacings will be conducted in our future research.}.

\subsubsection{\textbf{Numerical Results}}

\begin{figure}[!t]\centering
	\includegraphics[width=0.4\textwidth]{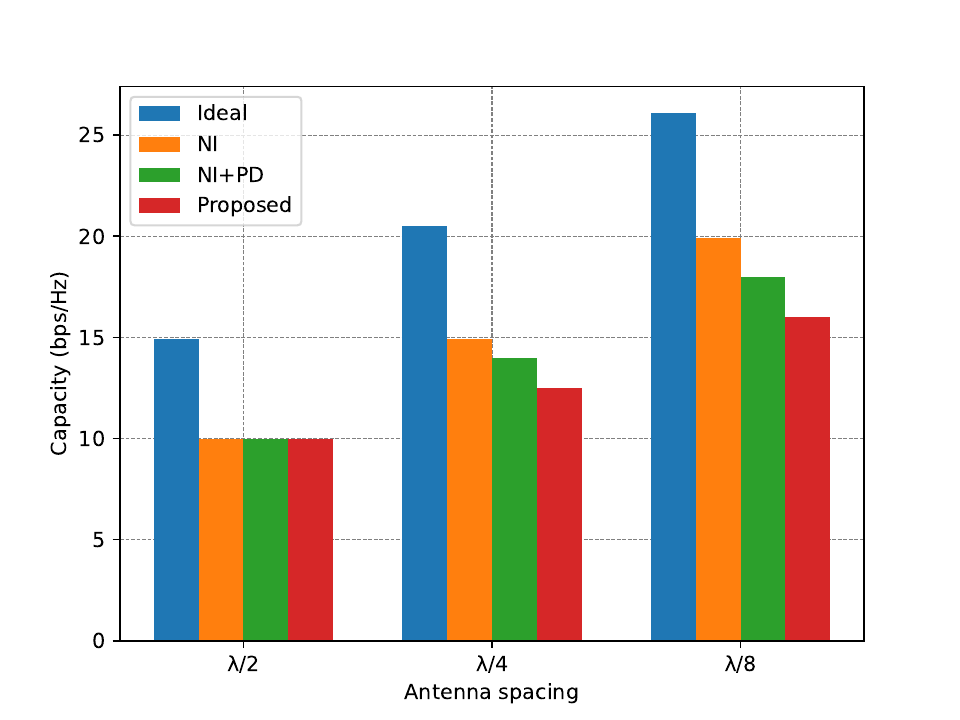}
	\caption{Channel capacity with different antenna spacings based on the simulated channel.}
	\label{fig1_holo_cap_simu}
\end{figure}

Fig.~\ref{fig1_holo_cap_simu} demonstrates the channel ergodic capacity with different antenna spacing distances based on the channel model in~\ref{equ-chn-pol-mat}.
In addition to the proposed channel model, three benchmark schemes are considered to evaluate the non-ideal factors as follows: 1) Ideal: isotropic scattering environment, no pattern distortion (PD) and antenna efficiency loss (EL) due to mutual coupling; 2) NI: non-isotropic scattering environment, no PD and antenna EL due to mutual coupling; 3) NI+PD: non-isotropic scattering environment, embedded PD but no antenna EL due to mutual coupling; 4) Proposed: non-isotropic scattering environment, embedded PD and 80\% antenna efficiency due to mutual coupling.

First, one can observe that the capacity increases as the antenna spacing distance decreases; however, the performance gain diminishes when non-ideal channel factors are considered. This phenomenon is due to the fact that increasing the number of antenna elements introduces additional multiplexing and diversity gains.
Second, it is shown that with the same antenna spacing, the capacity decreases with the increase of non-ideal factors. This is expected because the angular spread in the isotropic environment is significantly larger than in the non-isotropic environment, providing more spatial DoFs. The array and multiplexing gains achieved by more antenna elements are negated by the reduction in antenna efficiency\footnote{The present model extends the 3GPP TR 38.901 channel model with near-field spherical wavefront and spatial non-stationarity, while the gap between deterministic electromagnetic principles and statistical channel characterization will be addressed through hybrid models combining Green's function-based physics with data-driven stochastic corrections in our future work.}.

\begin{figure}[!t]\centering
	\includegraphics[width=0.4\textwidth]{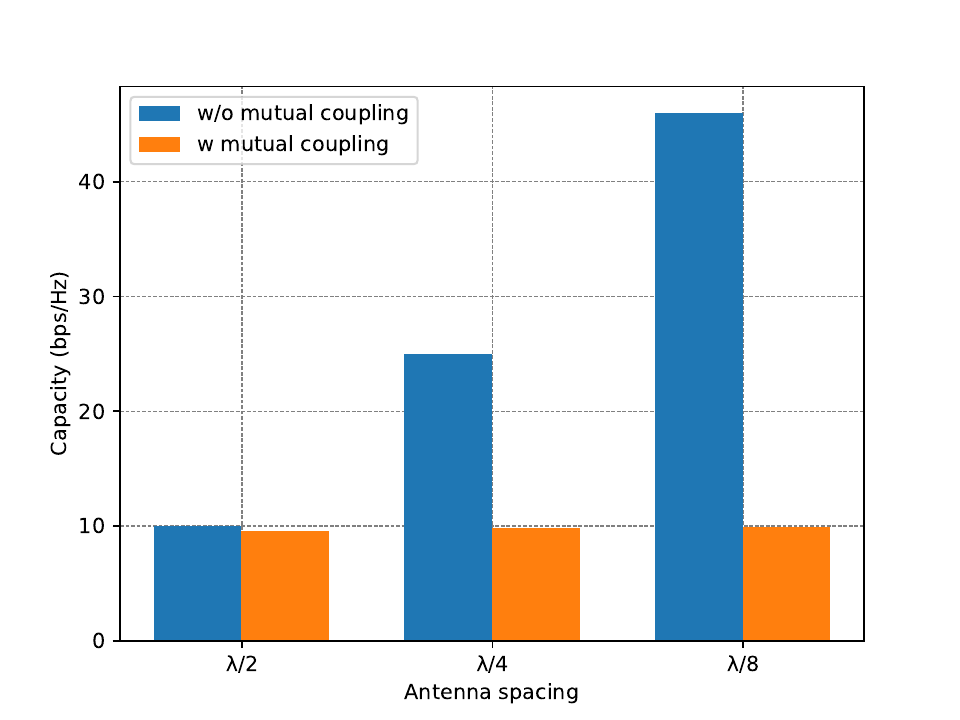}
	\caption{Channel capacity with different antenna spacings based on the measured channel.}
	\label{fig2_holo_cap_meas}
\end{figure}

Fig.~\ref{fig2_holo_cap_meas} evaluates the channel ergodic capacity with different antenna spacing distances based on the measured channel in an indoor environment specified before.
The equal power allocation strategy is adopted to evaluate the channel capacity performance.
It can be observed that, when mutual coupling effects are neglected, the capacity increases substantially with decreasing antenna element spacing, with gains reaching up to 450\%. However, the reduction in antenna efficiency due to mutual coupling significantly diminishes this capacity gain, posing a critical challenge for practical densely spaced MIMO systems. From an angular spectrum perspective, the theoretical capacity gain under ignored mutual coupling originates from accessing additional evanescent wave modes through spatial oversampling - these exponentially decaying components, when ideally transduced, provide extra degrees of freedom for MIMO operation. While mutual coupling severely diminishes the capacity gain under conventional array designs, this result highlights an essential research direction: developing antenna decoupling techniques to preserve the theoretical advantages of spatial oversampling. Mutual coupling effects need not be fundamental limitations, but rather engineering challenges to overcome for next-generation ultra-dense arrays.

\subsection{Near Field Communications} \label{sec-NFC}

\begin{figure*}[hb]
	\hrulefill
	\begin{equation}
		\begin{aligned}
			\mathbf{H}_{u,s}^{\text{LOS}}\left(t\right)\!= \!
			\left[\begin{matrix}F_{u,\theta}^{\text{Rx}}\left(\theta_{u}^{\text{Rx}},\varphi_{u}^{\text{Rx}}\right)\\F_{u,\phi}^{\text{Rx}}\left(\theta_{u}^{\text{Rx}},\varphi_{u}^{\text{Rx}}\right)\\\end{matrix}\right]^T\!\!	\left[\begin{matrix}1&0\\0&-1\\\end{matrix}\right]\!\!
			\left[\begin{matrix}F_{s,\theta}^{\text{Tx}}\left(\theta_{s}^{\text{Tx}},\varphi_{s}^{\text{Tx}}\right)\\F_{s,\phi}^{\text{Tx}}\left(\theta_{s}^{\text{Tx}},\varphi_{s}^{\text{Tx}}\right)\\\end{matrix}\right]
			\!\exp\!\left(\!\!-j2\pi\frac{d_{0,0}}{\lambda_0}\!\right) \!\exp\!\left(\!j2\pi\frac{d_{0,0}\!-\!d_{u,s}}{\lambda_0}\right) \!\exp\!\left(\!j2\pi\frac{{({\hat{r}^{\text{Rx}}_{u,s}})^T\!\cdot\!\bar{v}}}{\lambda_0}t\!\right).
		\end{aligned}\label{eq_los}
	\end{equation}
	\begin{equation}
		\begin{aligned}
			\mathbf{H}_{u,s,n,m}^{\text{NLOS}}\left(t\right)&=\sqrt{\frac{P_n}{M}}\left[\begin{matrix}F_{u,\theta}^{\text{Rx}}\left(\theta_{n,m,u}^{\text{Rx}},\varphi_{n,m,u}^{\text{Rx}}\right)\\F_{u,\varphi}^{\text{Rx}}\left(\theta_{n,m,u}^{\text{Rx}},\varphi_{n,m,u}^{\text{Rx}}\right)\\\end{matrix}\right]^T
			\left[\begin{matrix}\exp{\left(j\mathrm{\Phi}_{n,m}^{\theta\theta}\right)}&\sqrt{{\kappa_{n,m}}^{-1}}\exp{\left(j\mathrm{\Phi}_{n,m}^{\theta\varphi}\right)}\\\sqrt{{\kappa_{n,m}}^{-1}}\exp{\left(j\mathrm{\Phi}_{n,m}^{\varphi\theta}\right)}&\exp{\left(j\mathrm{\Phi}_{n,m}^{\varphi\varphi}\right)}\\\end{matrix}\right]
			\left[\begin{matrix}F_{s,\theta}^{\text{Tx}}\left(\theta_{n,m,s}^{\text{Tx}},\varphi_{n,m,s}^{\text{Tx}}\right)\\F_{s,\varphi}^{\text{Tx}}\left(\theta_{n,m,s}^{\text{Tx}},\varphi_{n,m,s}^{\text{Tx}}\right)\\\end{matrix}\right] \\
			&\exp{\left(j2\pi\frac{d_{n,m,0}^{\text{Rx}}-d_{n,m,u}^{\text{Rx}}}{\lambda_0}\right)}
			\,\exp{\left(j2\pi\frac{d_{n,m,0}^{\text{Tx}}-d_{n,m,s}^{\text{Tx}}}{\lambda_0}\right)}
			\exp\left(j2\pi\frac{{({\hat{r}^{\text{Rx}}_{u,n,m}})^T\cdot\bar{v}}}{\lambda_0}t\right).
		\end{aligned}\label{eq_nlos}
	\end{equation}
\end{figure*}

In this subsection, we propose an extended channel model by considering the spherical wave in near-field communications based on the 3GPP TR 38.901 channel model \cite{3GPP38901}. 
Moreover, the visible probability is utilized to model the spatial non-stationary effect. Both simulation and experimental results are provided to validate our proposed channel model.
\subsubsection{\textbf{Channel Model}}

In the existing 3GPP TR 38.901 channel model, the planar-wave assumption is adopted. However, this assumption is invalid in near-field communication scenarios, where the spherical wave propagation characteristics cannot be neglected. 
To characterize the spherical wave, we adjust the angle and distance of each ray observed at each antenna element, taking into account the specific locations of the Tx and Rx, and/or scatterers. 
This extension enables the characterization of spherical wavefronts, providing a more precise representation of signal propagation in near-field environments.

At Tx and Rx, $N_T$ and $N_R$ antenna elements are deployed, respectively, indexed by $s\in\left\{0,...,N_T-1\right\}$ and $u\in\left\{0,...,N_R-1\right\}$. 
Both line-of-sight (LOS) and non-line-of-sight (NLOS) paths are considered with the $0$-th antenna element being the reference element. 
Let $d_{u,s}$ denote the distance between the $u$-th antenna element at the Rx and the $s$-th antenna element at the Tx, which can be utilized to model the phase difference between different elements for the LoS path. The zenith and azimuth angles of arrival/departure for each Rx/Tx element for the LOS path are denoted by $\theta_{u}^{\text{Rx}}$, $\varphi_{u}^{\text{Rx}}$, $\theta_{s}^{\text{Tx}}$, and $\varphi_{s}^{\text{Tx}}$, respectively. Consequently, the LOS component of the channel can be expressed as in~\eqref{eq_los}. The definitions of other parameters in~\eqref{eq_los} are consistent with those in the 3GPP TR 38.901 channel model, where $F_{u,\theta}^{\text{Rx}}$, $F_{u,\varphi}^{\text{Rx}}$, $F_{s,\theta}^{\text{Tx}}$, and $F_{s,\varphi}^{\text{Tx}}$ represent the field patterns of the $u$-th/$s$-th antenna element at the Rx/Tx side in the vertical and horizontal directions, respectively. The spherical unit vector from the $u$-th Rx element to the $s$-th Tx element is denoted by $\hat{r}^{\text{Rx}}_{u,s}$. $\bar{v}$ represents the velocity vector of the user terminal, and $\lambda_0$ denotes the carrier wavelength.

\begin{figure}[!t]
	\centering
	\includegraphics[width=0.5\textwidth]{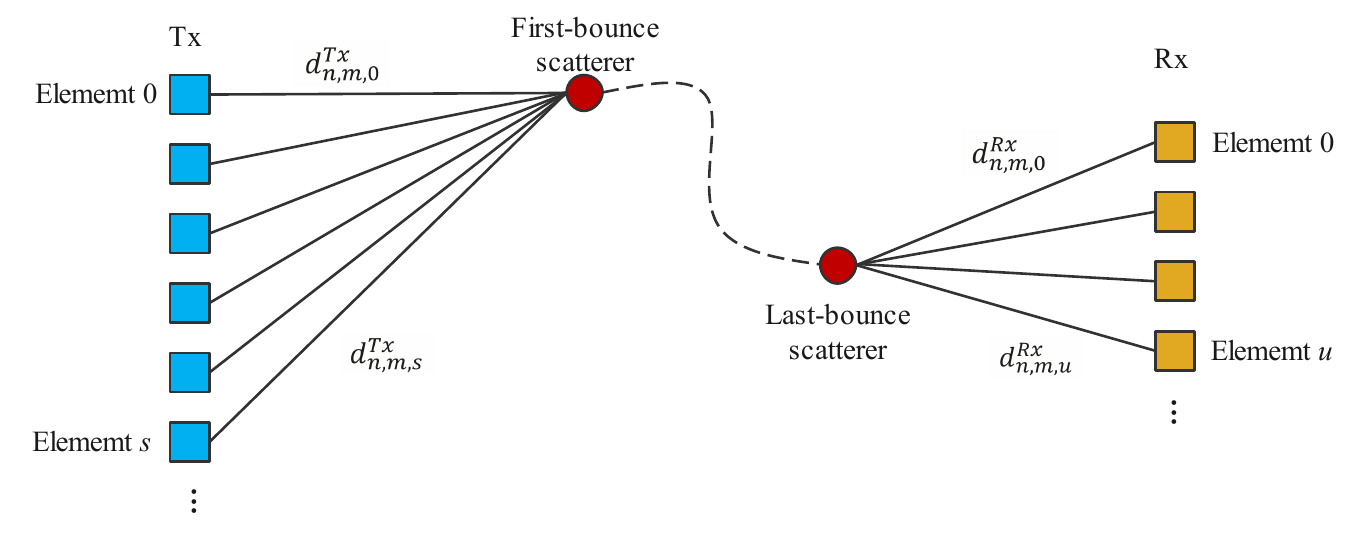}
	\caption{Illustration of spherical-wavefront model methodology for NLOS Path.}
	\label{fig_NLoS}
\end{figure}

Next, for the NLOS component of the channel, the delay and angle of a ray generated according to the existing channel model can be considered as the observed delay and angle at the reference antenna element. These parameters can be utilized to locate the first- and last-bounce scatterers, as illustrated in Fig.~\ref{fig_NLoS}. Based on the determined scatterer locations, the antenna gain and steering vector in the channel coefficient equation should be revised, similar to the approach for the LOS path, as shown in~\eqref{eq_nlos}. For the $m$-th ray in the $n$-th cluster, $d_{n,m,u}^{\text{Rx}}$ denotes the distance between the last-bounce scatterer and the $u$-th Rx antenna element, while $d_{n,m,s}^{\text{Tx}}$ represents the distance between the first-bounce scatterer and the $s$-th Tx antenna element. The zenith and azimuth angles of arrival/departure for each Rx/Tx element for the NLOS ray are denoted by $\theta_{n,m,u}^{\text{Rx}}$, $\varphi_{n,m,u}^{\text{Rx}}$, $\theta_{n,m,s}^{\text{Tx}}$, and $\varphi_{n,m,s}^{\text{Tx}}$, respectively. The second $2\times2$ matrix represents the cross-polarization characteristics, and $\hat{r}^{\text{Rx}}_{u,n,m}$ denotes the spherical unit vector from the $u$-th Rx element to the first-bounce of the $m$-th ray in the $n$-th cluster. Additionally, $P_n$ represents the power of the $n$-th cluster, which contains $M$ rays.

Thus, the channel impulse response, obtained by superimposing the LOS and NLOS path components and scaling both terms according to a K-factor $K_R$, can be expressed as  
\begin{equation}  
	\begin{aligned}  
		&\mathbf H_{u,s}(\tau,t) =\sqrt{\frac{K_R}{K_R+1}}\mathbf H_{u,s}^\text{LOS}(t)\delta(\tau-\tau_1) \\
		& +\sqrt{\frac{1}{K_R+1}}\sum_{n,m}{\mathbf H_{u,s,n,m}^\text{NLOS}(t)\delta(\tau-\tau_{n,m,0,0})},
	\end{aligned}  
\end{equation}  
where $\tau_1$ and $\tau_{n,m,0,0}$ represent the propagation delays from the Tx reference antenna element to the Rx reference antenna element through the LOS path and the NLOS ray $(n, m)$, respectively. This implies that the delay difference between antenna elements at the Tx/Rx sides is negligible.

Furthermore, a visible probability is considered to capture the spatial non-stationary effect. For the $n$-th cluster, the visible probability of each antenna element can be modeled using an exponential distribution that is negatively related to the cluster power $P_n$, as expressed by the following equation:  
\begin{equation}  
	V_n = A\cdot e^{\frac{-\max(P_n)-P_n}{\lambda}}+ B +\delta,
\end{equation}  
where $A$ and $\lambda$ are the parameters of the exponential distribution, $B$ represents the lower bound of the visible probability, and $\delta\sim \mathbf N(0,\xi^2)$ follows the Gaussian normal distribution. This indicates that weaker rays, characterized by lower power, are more susceptible to spatial non-stationarity.
Then, a visible-probability-related power attenuation factor for each ray, denoted as $\alpha_{s,n,m}$, is introduced to reflect the impact of spatial non-stationarity. This factor is given by 
\begin{equation}
	\alpha_{s,n,m}=\frac{1}{1+\exp{\left(\Delta d \cdot C\right) }}, 
\end{equation} 
where $\Delta d = (\frac{d_{n,m,s}^\text{Tx}-d_{n,m,s}^{\text{Tx,min}}}{d_{n,m,s}^{\text{Tx,max}}-d_{n,m,s}^{\text{Tx,min}}}-V_n)$ considers the distance between each cluster and Tx antenna element, and $C$ represents the roll-off coefficient between the visible and invisible regions. Subsequently, the power attenuation factor is applied to the corresponding channel coefficient of each path component, as follows:  
\begin{align}  
	\hat{\mathbf H}_{u,s}^\text{LOS}(t) &= \alpha_{s,0,0}\cdot \mathbf H_{u,s}^\text{LOS}(t), \\  
	\hat{\mathbf H}_{u,s,n,m}^\text{NLOS} &= \alpha_{s,n,m}\cdot \mathbf H_{u,s,n,m}^\text{NLOS}(t).  
\end{align}  

Consequently, both spherical-wave characteristics and spatial non-stationarity effects have been incorporated into the proposed extended channel model.

\subsubsection{\textbf{Simulation and Measurement Setup}}
In the simulation, we consider two scenarios with central frequencies of 6.7 GHz and 15 GHz. The dimensions of the antenna arrays at the BS are set to $\left\{0.33 \text{m}, 1.5 \text{m}\right\}$ for the 6.7 GHz case and $\left\{0.32 \text{m}, 1.36 \text{m}\right\}$ for the 15 GHz case. The Rayleigh distances for the near-field can be derived as 104 m and 196 m, respectively, with array apertures of $D=1.53 \text{ m}$ and $D=1.4 \text{ m}$. Notably, the 15 GHz scenario exhibits a larger near-field region. Furthermore, one hundred UEs are randomly distributed at least 35 m away from the BS, with $80\%$ located indoors and the remaining outdoors, experiencing random LOS with NLOS channel conditions. The impact of spherical-wavefront modeling is investigated by analyzing the spatial channel difference of planar-wavefront and spherical-wavefront assumptions. This difference is characterized by their correlations as $\rho=\frac{\left|\mathbf h_\text{PWA}^H\mathbf h_\text{SWA}\right|}{\left|\mathbf h_\text{PWA}\right|\left|\mathbf h_\text{SWA}\right|}$, where $h_\text{PWA}$ and $h_\text{SWA}$ represent the vectorized spatial channel metrics under planar-wavefront and spherical-wavefront assumptions, respectively.

\begin{figure}[!t]
	\centering
	\includegraphics[width=0.4\textwidth]{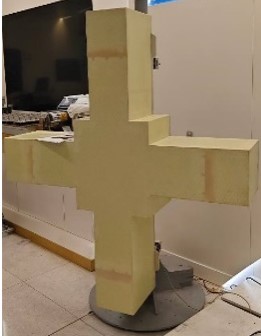}
	\caption{Illustration of experimental antenna arrays at the base station and the UEs.}
	\label{fig_antenna}
\end{figure}

Furthermore, to validate our proposed channel model, an empirical outdoor near-field channel measurement was conducted. The channel data collected from this experiment was compared with the simulated channel from the proposed channel model. The schematic diagram of the measurement equipment is shown in Fig.~\ref{fig_antenna}. In the experimental setup, a central frequency of 6.7 GHz was used, and a cross-shaped antenna array with dimensions of $1.4\text{m} \times 1.4\text{m}$ was employed at the BS, with 64 antenna elements in both the horizontal and vertical directions. At the UE side, a standard antenna array consisting of 4 antennas with a spacing of half wavelength was deployed. The distance between the Tx and Rx was set to 20 m. We examined the phase deviation of the channel coefficients caused by spherical wave propagation and collected the raw visibility ratio of the Tx antenna elements to valuate the non-stationarity effect.

\subsubsection{\textbf{Numerical Results}}
\begin{figure}[!t]
	\centering
	\subfloat[\label{fig_corr_u6g}]{
		\includegraphics[width=0.4\textwidth]{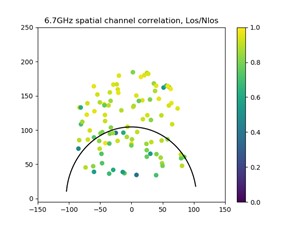}}
	\quad	
	\subfloat[\label{fig_corr_15g}]{
		\includegraphics[width=0.4\textwidth]{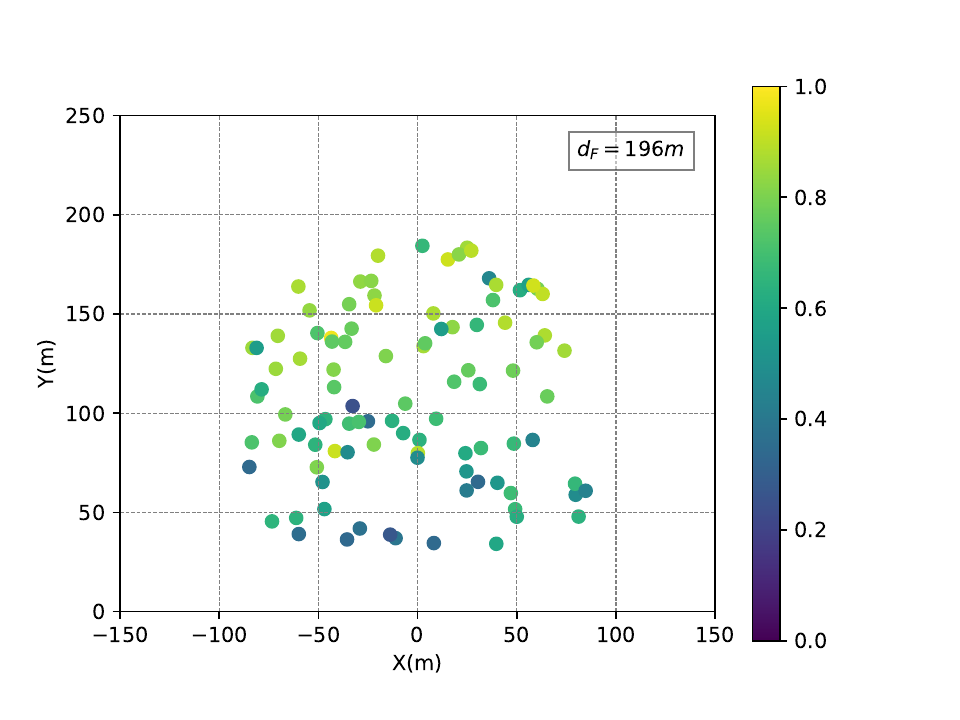}}
	\caption{Spatial channel correlation between the near-field model and far-field model for 6.7~GHz and 15~GHz scenarios.}
	\label{fig_corr} 
\end{figure}

Fig.~\ref{fig_corr} presents the spatial channel correlations of the simulated channel under planar- and spherical-wave assumptions. As evident from Fig.~\ref{fig_corr}, UEs located closer to the BS exhibit significantly lower channel correlation, indicated by darker dots, which signify a greater disparity between the channel coefficients under planar and spherical-wavefront assumptions. In the 15 GHz scenario, all UEs within the sector are within the Rayleigh distance, resulting in lower channel correlations compared to the 6.7 GHz scenario.

\begin{figure}[!t]
	\centering
	\includegraphics[width=0.4\textwidth]{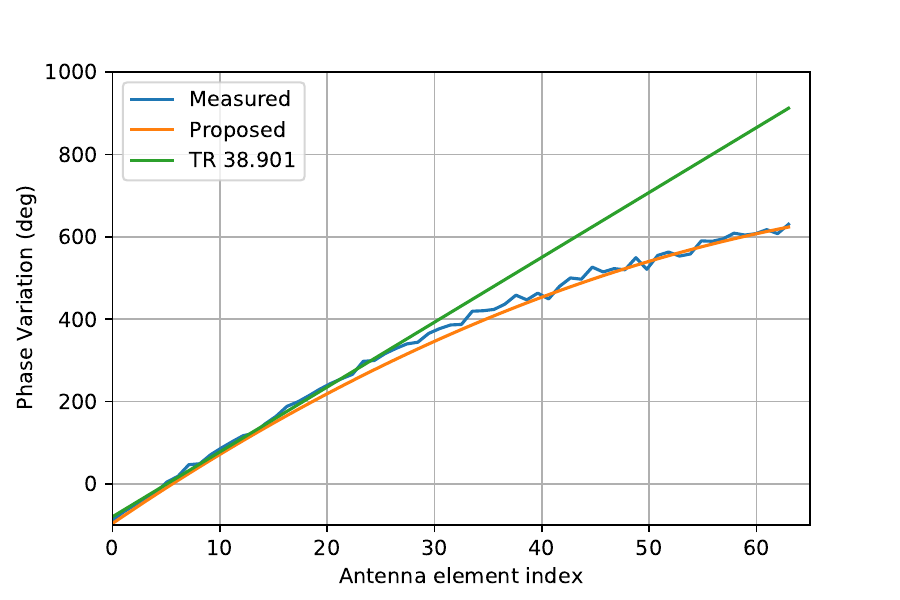}
	\caption{Comparison of LOS path phase variation between the near-field model and the far-field model.}
	\label{fig_ex_sphere}
\end{figure}

\begin{figure}[!t]
	\centering
	\includegraphics[width=0.4\textwidth]{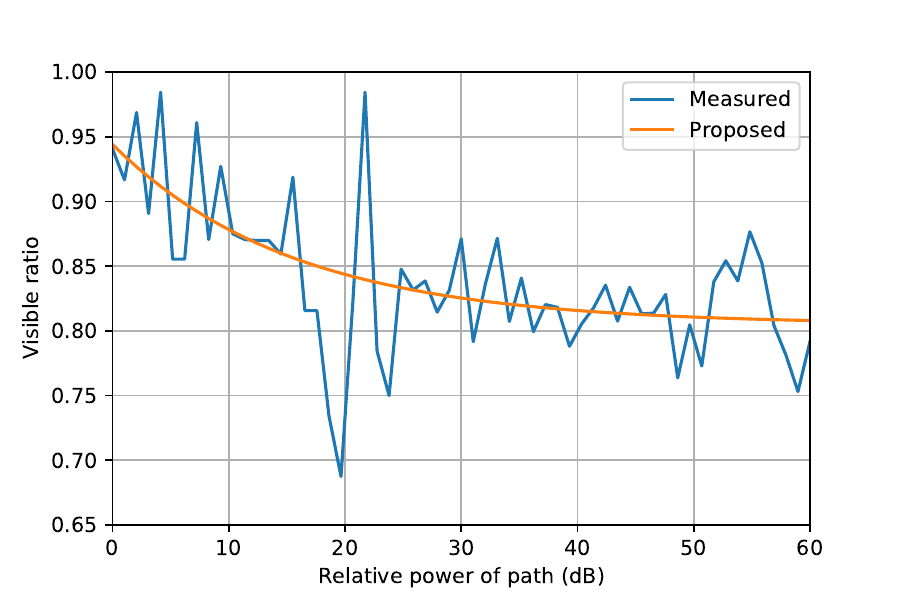}
	\caption{Comparison of visible probability of the measured results and the proposed model.}
	\label{fig_ex_sns}
\end{figure}

Next, we present the experimental analysis results based on the measured channel. As illustrated in Fig.~\ref{fig_ex_sphere}, there is a significant gap between the measured channel and the 3GPP TR 38.901 channel with a planar-wave assumption. Notably, the nonlinear phase variation of the LOS path observed in actual near-field propagation, represented by the blue curve, closely resembles that of our proposed channel model. This demonstrates that our proposed model can more accurately capture the phase characteristics in the near field. Subsequently, the raw visibility ratio of the Tx antenna elements for the non-stationarity effect is presented in Fig.~\ref{fig_ex_sns}. Our findings indicate that our proposed exponential-distribution visibility probability model can accurately fit the measured visibility ratio of the Tx elements.

\subsection{Tri-Polarized MIMO} \label{sec-tri}

In this subsection, we propose a joint uplink-downlink channel estimation method tailored for tri-polarized MIMO systems. Unlike conventional dual-polarized antennas limited to horizontal (H) and vertical (V) polarizations, tri-polarized MIMO introduces a third orthogonal polarization (z-axis) to address energy imbalance issues caused by polarization mismatch in NLOS scenarios \cite{4570210}. Specifically, the z-polarized component enhances spatial diversity by capturing signals from elevated scattering paths (e.g., reflections from buildings in urban environments), which are poorly resolved by H/V-polarized waves. This additional polarization direction improves channel capacity and robustness while maintaining backward compatibility with dual-polarized systems. We validate the proposed method through system-level simulations incorporating tri-polarized antenna radiation patterns and the 3GPP TR 38.901 channel model under practical communication scenarios.

\begin{figure*}[!t]
	\centering
	\includegraphics[width=0.9\textwidth]{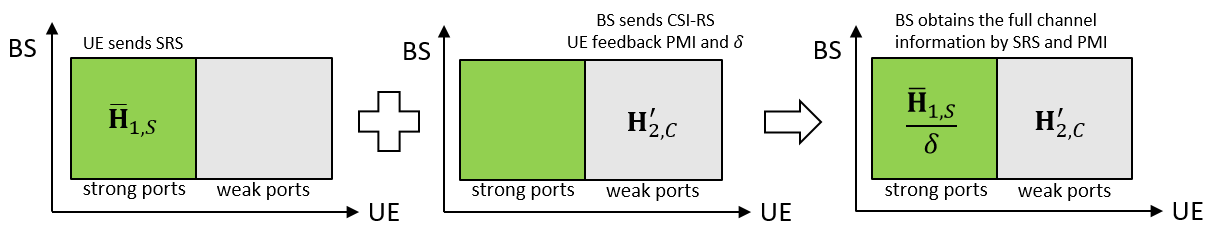}
	\caption{Illustration of integrated channel estimation method for tri-polarized MIMO.}
	\label{f_tri_scheme}
\end{figure*} 

\subsubsection{\textbf{Channel model}}
We consider a tri-polarized wireless communication system, where $N_\mathrm{S}=N^\mathrm{x}_\mathrm{S} + N^\mathrm{y}_\mathrm{S} + N^\mathrm{z}_\mathrm{S}$ and $N_\mathrm{R}=N^\mathrm{x}_\mathrm{R} + N^\mathrm{y}_\mathrm{R} + N^\mathrm{z}_\mathrm{R}$ antennas are employed at the Tx and Rx, respectively, with $N^{i}_{S}$ and $N^{i}_{R}$ denoting the number of $i$-th polarization antennas at Tx and Rx. The received signal is then given as
\begin{equation}
	\mathrm{\textbf{y = Hs + n}},
\end{equation}
with
\begin{equation}
	\mathrm{\textbf{H}}= 
	\begin{bmatrix}
		\mathrm{\textbf{H}}_{xx} \;\; \mathrm{\textbf{H}}_{xy} \;\; \mathrm{\textbf{H}}_{xz}   \\
		\mathrm{\textbf{H}}_{yx} \;\;  \mathrm{\textbf{H}}_{yy} \;\; \mathrm{\textbf{H}}_{yz}  \\
		\mathrm{\textbf{H}}_{zx} \;\;  \mathrm{\textbf{H}}_{zy} \;\; \mathrm{\textbf{H}}_{zz} 
	\end{bmatrix},
\end{equation}
where $\mathrm{\textbf{H}} \in {\mathbb{C}^{N_R \times N_S}}$ is the tri-polarized channel, and $\mathrm{\textbf{H}}_{i,j}  \in {\mathbb{C}^{N_R^i \times N_S^j}}$ represents the channel between the antenna ports of $i$-th polarization at Rx and that of $j$-th polarization at Tx.

\subsubsection{\textbf{Proposed Method}}
\begin{figure}[!t]
	\centering
	\subfloat[\label{fig_tri_pol}]{
		\includegraphics[width=0.8\linewidth]{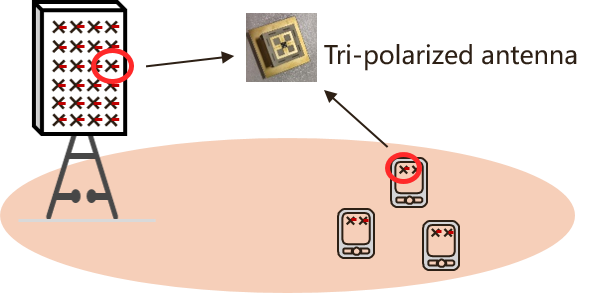}}
	\quad \\
	\subfloat[\label{fig_XPO}]{
		\includegraphics[width=0.45\linewidth]{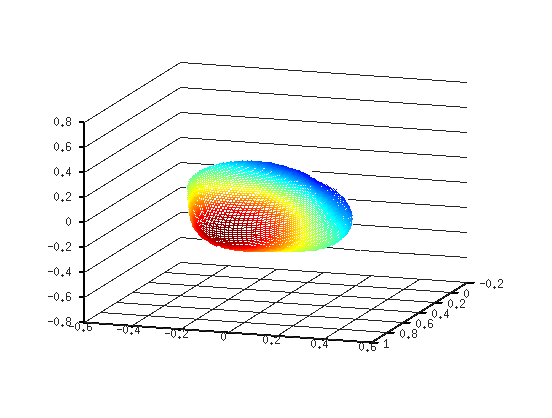}}
	\subfloat[\label{fig_TRI}]{
		\includegraphics[width=0.45\linewidth]{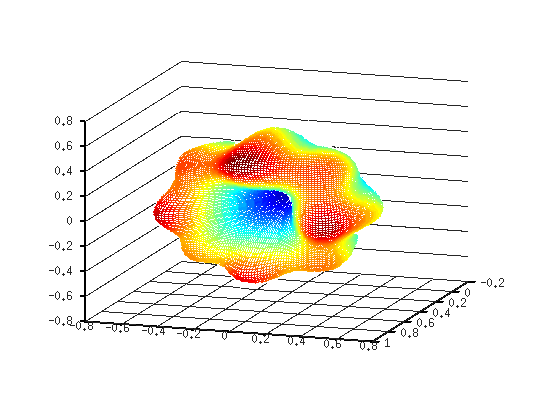}}
	\caption{Illustration of the antenna radiation patterns at UEs. (a) Test scenario. (b) The first two polarizations. (c) The third polarization.}
	\label{fig_radiation}
\end{figure}

We consider a communication system operating in time division duplexing (TDD) mode.
The directional radiation pattern employed in the practical system is considered to enhance the channel gain in a specific direction. Note that the third-polarized antenna can augment the existing dual-polarized radiation pattern, as illustrated in Fig.~\ref{fig_radiation}, which facilitates more efficient full-space reception.

Downlink channel estimation can be conducted by transmitting downlink pilots, based on which the Rx estimates the channel and feeds back quantized channel information via the uplink channel. Alternatively, the downlink channel can be obtained by measuring the uplink channel and leveraging the inherent reciprocity between the uplink and downlink channels in the TDD communication system.
However, the channel estimation accuracy for some antenna ports deteriorates due to variations in antenna gain among different polarizations. 
To mitigate this issue, a joint uplink-downlink channel estimation method is proposed. As shown in Fig.~\ref{f_tri_scheme}, the main idea of the proposed method involves measuring the channel for port groups with high antenna gain using the uplink channel estimation method, while measuring the channel for port groups with low antenna gain through downlink channel estimation.
The specific steps are as follows, illustrated with single-user MIMO system as an example.
\\
\noindent - \textit{{Step 1: Port grouping}}: Based on the receive power on Rx's each antenna port, the antenna ports are categorized into two distinct groups: one for ports with high antenna gain (marked as G1), and another for those with low antenna gain (marked as G2).

\noindent - {\textit{Step 2: Uplink channel estimation for G1}}:
The Rx sends pilots through the ports in G1, such that Tx can estimate the channel for corresponding ports, denoted by $\mathrm{\textbf{H}_{1,U}}$, by leveraging the channel reciprocity.

\noindent - {\textit{Step 3: Downlink channel measurement for G2}}:
The Tx sends pilots and then the Rx measures the full-dimension channel $\mathrm{\textbf{H}_\text D}$, which consists of two components corresponding G1 and G2, respectively, denoted by $\textbf{H}_{1,\text D}$ and $\textbf{H}_{2,\text D}$ with $\mathrm{\textbf{H}_\text D = [\textbf{H}_{1,D}, \textbf{H}_{2,D}]}$.
Next, the estimated channel for G2 is normalized and then fed back to Tx, given by
\begin{equation}
	\mathrm{\textbf{H}^{'}_{2,D}} = \frac{\mathrm{\textbf{H}_{2,D}}}{\rho_2 e^{j\omega_2}},
\end{equation}
where $\rho_2$ and $\omega_2$ represents the normalization amplitude and phase, respectively. Similarly, estimated channel for G1 can be normalized with a normalization amplitude and phase of $\rho_1$ and $\omega_1$.

\noindent - {\textit{Step 4: Port combining reference value feedback}}:
The estimated channel of antenna ports in G2 by downlink measurement cannot be directly combined with that in G1 due to the unknown combining reference value.
Therefore, the Rx needs to feedback the channel combining reference value, denoted by $\mathrm{\delta}$, from the full channel information $\mathrm{\textbf{H}_D}$, and is given by
\begin{equation}
	\mathrm{\delta} = \frac{\rho_1 e^{j\omega_1}}{\rho_2 e^{j\omega_2}}.
\end{equation}

\noindent - {\textit{Step 5: Joint channel estimation}}:
The estimated channel of antenna ports in G1, $\mathrm{\textbf{H}_{1,U}}$, is first normalized as $\mathrm{\bar{\textbf{H}}_{1,U}}$ and further adjusted with the channel combining reference value for channel combining as
\begin{equation}
	\mathrm{\textbf{H}^{'}_{1,U}} = \frac{\mathrm{\bar{\textbf{H}}}_{1,U}}{\delta}.
\end{equation}
As such, the channel for all antenna ports can be obtained as 
\begin{equation}
	\mathrm{\textbf{H}} = [\mathrm{\textbf{H}^{'}_{1,U}},\mathrm{\textbf{H}^{'}_{2,D}} ].
\end{equation}

\subsubsection{\textbf{Performance Evaluation}}
We conduct system-level simulation to evaluate the efficiency of our proposed joint uplink-downlink channel estimation method by taking into account real-world scenarios with tri-polarized MIMO.
It is assumed that 50 UEs are randomly distributed in a standard cell with three sectors.
The BS and each UE is equipped with 256 and 12 antenna ports respectively. The directional radiation patterns of the tri-polarized antennas used are illustrated in Fig.~\ref{fig_radiation}. Other simulation assumptions are detailed in Table~\ref{tab_par}.

\begin{table}[!t]
	\renewcommand{\arraystretch}{1.2}
	\centering
	\caption{Assumption for Tri-Polarized Channel Estimation}
	\label{tab_par}
	\begin{tabular}{cc}
		\toprule[1.2pt]
		\textbf{Simulation Parameter}  & \textbf{Value} \\ 
		\midrule
		\textbf{Scenario}  & 3 cells, 50 UE per cell \\
		\midrule
		\textbf{Center frequency}  & 6.7GHz \\
		\midrule
		\textbf{Channel model}  & 3GPP TR 38.901 \\
		\midrule
		\textbf{BS antenna setup}  & 256 ports: 0.33 m(H)x1.5 m(V)\\
		\midrule
		\textbf{UE antenna setup}  & - 8 ports: (M, N, P) = (2, 4, 2)\\
		&  - 12 ports: (M, N, P) = (2, 6, 3)\\
		&  (dH, dV) = (0.5$\lambda$, 0.5$\lambda$)\\
		\midrule
		\textbf{BS power and height}  & 39.64 dBm and 25 m\\
		\midrule
		\textbf{Traffic model}  & Full buffer\\
		\midrule
		\textbf{UE distribution}  & 80\% indoor, 20\% outdoor \\
		\bottomrule[1.2pt]
	\end{tabular}
\end{table}

\begin{figure}[!t]
	\centering
	\includegraphics[width=0.4\textwidth]{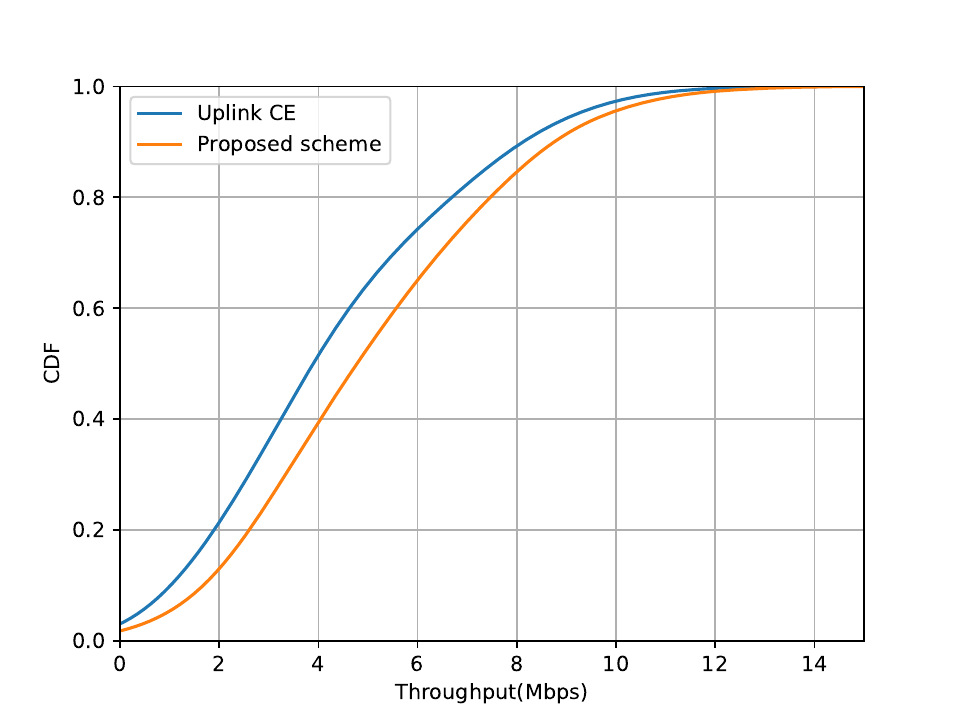}
	\caption{Comparison of average capacity for different channel estimation schemes for tri-polarized UEs.}
	\label{f_tri}
\end{figure}

Fig.~\ref{f_tri} illustrates the cumulative distribution function (CDF) curves of the capacity, comparing a benchmark utilizing uplink channel estimation with the proposed method. It is observable that the capacity of the proposed method consistently surpasses the benchmark across all intervals. Furthermore, it is demonstrated that the average capacity attained by the proposed method is 29\% higher than that of the benchmark.
This is expected because our proposed method efficiently compensate for the imbalance in receive power among different antenna ports.

\section{Future Research Directions}

In previous sections, we present the recent advances of the techniques in the field of EIT, and show their great potential in future 5G-Advanced era by integration of the physical characteristics, theoretical foundations and enabling technologies of new paradigm of MIMO communications. 
It should be recognized that this area is still in its early stages, presenting numerous research challenges that require careful attention, as well as new opportunities for practical application.
In the following, we identify and discuss some of the most critical research challenges of the EIT applications from the industrial perspective.

\subsection{Densely Spaced MIMO}

The core challenge lies in overcoming hardware imperfections, signal loss, mutual coupling, and environmental sensitivities to achieve efficient, scalable densely spaced MIMO systems for high-performance multi-user service. Edge elements in finite arrays experience less coupling, and dense irregular arrays may enhance central performance. Densely spaced MIMO channel estimation requires cost-efficient methods to accurately estimate channels with lower pilot overhead, without sacrificing performance given massive antenna elements and complex generic or sparse channel models. Efficient hardware design, low-cost channel estimation, and beamforming algorithms considering non-negligible mutual coupling warrant investigation.

\subsection{Near-field Communications}

One significant challenge in near-field communications is defining the boundary of near and far fields in practical propagation environments where LOS and NLOS paths coexist, such that a suitable communication paradigm can be selected to leverage the additional DoFs in near field.
Another area of focus is the development of efficient methods for multi-path near-field beam training and beam tracking, particularly in scenarios involving multiple NLOS paths and high mobility. 
Lastly, it is worthwhile to explore strategies for expanding the near-field range by altering the MIMO architecture and/or the propagation environment to construct a  favorable near-field propagation environment.

\subsection{Tri-polarized MIMO}

Firstly, maintaining high cross-polarization isolation is essential to prevent signal leakage in tri-polarized systems, challenging manufacturing processes compared to dual-polarization designs. Secondly, significant received energy differences between polarizations require joint uplink-downlink channel estimation schemes, while inconsistent measurement dimensions necessitate tailored pilot designs in time-frequency domains for tri-polarized scenarios. Lastly, existing tri-polarized MIMO research focuses on reception, while transmit-end performance remains underexplored.

\subsection{Other Emerging Paradigms}

There is a need to refine theoretical analyses that bridge classical information theory, which assumes Gaussian distributions for signals and noise, with precise channel capacity characterizations given exact electromagnetic information about the wireless propagation environment.
Additionally, research should build upon EIT's frameworks to achieve precise modeling of non-ideal electromagnetic properties in current communication systems, aiming to optimize the utilization of existing DoFs. In parallel, there should be an emphasis on innovating MIMO architectures to explore and leverage new communication dimensions, thus pushing the boundaries of wireless communication capabilities.

\section{Conclusions}

In conclusion, this paper has highlighted the emerging research direction of EIT, which seeks to integrate Shannon-based information theory with physical consistency, particularly the electromagnetic properties of communication channels. We have proposed an EIT-based MIMO paradigm that bridges the gap between conventional MIMO systems and next-generation MIMO systems by introducing the concepts of EM precoding and EM combining. The exploration of typical applications based on EIT theory, such as densely spaced MIMO, near-field communications, and tri-polarized MIMO, has shown significant potential for improving spectral efficiency and providing more transmission degrees of freedom. While there are practical challenges, this paper emphasizes the importance of EIT in enhancing communication systems, aligning with Shannon theory while addressing the need for more physically accurate models.

\bibliographystyle{IEEEtran}
\bibliography{bib/bib_book,bib/bib_paper}

\vfill

\end{document}